\documentclass[journal]{IEEEtran}
\usepackage{amsmath,bm,amssymb}
\usepackage{cancel}
\usepackage{amsfonts} 
\usepackage{enumerate}
\usepackage{xcolor}
\usepackage{graphicx}
\usepackage{caption}
\usepackage{subcaption}
\usepackage{svg} %vector figures
\usepackage{algorithm}% http://ctan.org/pkg/algorithms
\newcommand{\Inputs}[1]{\State \textbf{Inputs:} #1}
\newcommand{\Outputs}[1]{\State \textbf{Outputs:} #1}

\usepackage{algpseudocode}% http://ctan.org/pkg/algorithmicx
\usepackage[normalem]{ulem}
\usepackage{makecell} %How to add a forced line break inside a table cell
\usepackage{nomencl} %nomenclature definition
\usepackage[utf8]{inputenc}
\usepackage{arydshln} %dashline in matrix
\usepackage{cuted}
\usepackage{textgreek} 
\usepackage{cite} % To convert citation [1],[2],[3],[4] to [1]-[4]
\usepackage{multirow} % Table \multirow
\usepackage{diagbox} % 
\usepackage{float}
\usepackage{anyfontsize} 
\usepackage{orcidlink}

\title{Detecting Cyber Attacks in Power System AGC Using a Drifted Ornstein-Uhlenbeck Process}

\begin{document}

\author{
Mingqiu Du~\orcidlink{0000-0002-6486-1067}, 
    Xiaozhe Wang$^{\ast}$~\orcidlink{0000-0002-1887-0990},~\IEEEmembership{Senior Member,~IEEE}, 
    Qinglai Guo~\orcidlink{0000-0003-1435-5796},~\IEEEmembership{Fellow,~IEEE}

\thanks
{This work is supported by Hydro-Qu\'{e}bec, the Institut de Valorisation des Donn\'{e}es (IVADO), MITACS under Grant IT27493, the Natural Sciences and Engineering Research Council of Canada (NSERC) under Alliance Grants ALLRP 566986-21, and Fonds de recherche du Qu\'{e}bec--Nature et technologies (FRQNT) under Grants FRQ-NT PR-256837 and 298827. The authors would also like to thank Dr. Marthe Kassouf for her role in initiating this collaboration. 

M. Du and X. Wang are with the Department of Electrical and Computer Engineering, McGill University, Montreal, QC H3A 0E9, Canada (e-mail: mingqiu.du@mail.mcgill.ca; xiaozhe.wang2@mcgill.ca).
Q. Guo is with the Department of Electrical Engineering, Tsinghua University, Beijing, 100084, China (guoqinglai@tsinghua.edu.cn).}
}
\maketitle
\begin{abstract}
The Automatic Generation Control (AGC) system, reliant on real-time measurements over communication networks, is susceptible to stealthy false data injection attacks (FDIAs), risking equipment damage and economic losses. We propose a robust FDIA detection method using maximum likelihood estimation (MLE) of a drifted multivariate Ornstein-Uhlenbeck (OU) process. Independent of load observability, in various cyberattack scenarios, the proposed FDIA detection method delivers accurate and rapid detection of sophisticated FDIAs, outperforming traditional unknown input observer (UIO) methods, which miss detections, and Long Short-Term Memory Autoencoder (LSTM-AE) approaches, which suffer from prolonged detection times.
\end{abstract}
%%%%%%%%%%%%%%
\begin{IEEEkeywords}
Automatic generation control, false data injection attack,  drifted multivariate Ornstein-Uhlenbeck process, cyber-physical security.
\end{IEEEkeywords}
%%%%%%%%%%%%%%
%====================== Nomenclature ======================
\section*{Nomenclature}

\def\nomlabelwidth{\IEEEsetlabelwidth{$\hat A_{sub},\hat{\bm{\mu}}_{sub},\hat\Sigma_{sub}$}}

\noindent\textit{Acronyms}
\begin{IEEEdescription}[\nomlabelwidth]
\item[AGC] Automatic Generation Control
\item[ACE] Area Control Error
\item[MLE] Maximum likelihood estimation
\item[LSTM-AE] Long Short-Term Memory Autoencoder
\item[PMU] Phasor measurement unit
\item[PDF] Probability density function
\item[FDIA] False data injection attack
\item[OU] Ornstein-Uhlenbeck
\item[UIO] Unknown input observer
\item[KF] Kalman filter
\item[CUSUM] Cumulative sum
\item[RTU] Remote terminal unit
\item[BDD] Bad data detection
\end{IEEEdescription}

\noindent\textit{State variables and stochastic quantities}
\begin{IEEEdescription}[\nomlabelwidth]
\item[$\Delta f_i$] Frequency deviation of area $i$
\item[$\Delta P_{tie_{i-j}}$] Tie-line power deviation from area $i$ to $j$
\item[$\Delta P_{m_i}$] Mechanical power deviation of area $i$
\item[$ACE_i$] Area control error of area $i$
\item[$\gamma_i$] Load-diffusion coefficient of area $i$
\item[$\bm{x}_t$] Continuous-time AGC state vector
\item[$\bm{x}_{sub}^{(k)}$] Measurable subsystem state vector
\item[$\bm{\mu}_t,\bm{\mu}$] OU equilibrium (drift mean) vector
\item[$\bm{\xi}_t$] Standard Gaussian white-noise vector
\item[$\tilde{\bm{x}}_t, \tilde{\bm{x}}^{(k)}$] Injected false-data signal
\item[$\Delta P_{ref_i}$] Power-reference deviation of area $i$
\item[$\Delta P_{g_i}$] Governor power deviation of area $i$
\item[$\Delta P_{L_i}$] Load deviation of area $i$
\item[$\mu_{L_i}$] Mean load deviation of area $i$
\item[$K_{L_i}$] Load mean-reversion coefficient of area $i$
\item[$\bm{x}^{(k)}$] Discrete-time state vector at sample $k$
\item[$\bm{x}_{oth}^{(k)}$] Unmeasured state subvector
\item[$\bm{\mu}_L$] Vector of mean load deviations
\item[$\bm{W}_t$] Wiener process
\item[$\Delta \tilde f_i$] False frequency injected in area $i$
\item[$\bm{v}^{(k)}$] Measurement noise vector at sample $k$
\item[$\hat{\bm{x}}^{(k|k - 1)}$] Predicted state estimate in KF
\item[$\hat{\bm{x}}^{(k|k)}$] Updated state estimate in KF
\item[$\tilde{\bm{y}}^{(k)}$] Measurement residual (innovation) vector
\end{IEEEdescription}

\noindent\textit{Matrices and operators}
\begin{IEEEdescription}[\nomlabelwidth]
\item[$A$] Continuous drift/state matrix
\item[$S$] OU diffusion-input matrix
\item[$\Sigma_{sub}$] Subsystem covariance matrix
\item[$V$] Tie-line incidence matrix
\item[$e^{A\Delta t}$] Discrete state-transition matrix
\item[$\hat A_{sub},\hat{\bm{\mu}}_{sub},\hat\Sigma_{sub}$] Online detection subsystem
\item[$A_{sub}$] Subsystem drift matrix
\item[$\Sigma$] Conditional covariance matrix
\item[$Q$] Measurable state selection
\item[$I$] Identity matrix
\item[$\hat A,\hat{\bm{\mu}},\hat\Sigma$] MLEs of $A$,$\bm{\mu}$,$\Sigma$
\item[$P^{(k|k - 1)}$] Predicted error covariance matrix in KF
\item[$P^{(k|k)}$] Updated error covariance matrix in KF
\item[$Q_{kf}, R_{kf}$] Process \& measurement noise covariances
\item[$S_{kf}^{(k)}$] Innovation covariance matrix in KF
\item[$K_{kf}^{(k)}$] Kalman gain matrix
\item[$A_2$] Unmeasured state dynamics matrix
\end{IEEEdescription}
\noindent \textit{Note:} $D = \mathrm{diag}[D_1,\dots,D_n]$, and similarly for $R, T_T, K_a, K_L, \Gamma, H, T_g, B$; $K_t = \mathrm{diag}[K_{tie_1},\dots,K_{tie_m}]$.

\noindent\textit{Physical and control parameters}
\begin{IEEEdescription}[\nomlabelwidth]
\item[$H_i$] Equivalent inertia of area $i$
\item[$R_i$] Speed regulation of area $i$
\item[$T_{T_i}$] Turbine time constant of area $i$
\item[$B_i$] Frequency-bias factor of area $i$
\item[$\Delta t$] Sampling interval
\item[$D_i$] Load-damping coeff. of area $i$ 
\item[$T_{g_i}$] Governor time constant of area $i$
\item[$K_{a_i}$] AGC integrator gain of area $i$
\item[$K_{tie_{i-j}}$] Sync. coeff. between areas $i, j$
\item[$M_{scale}$] Inertia multiplier
\item[$H_{i,scaled}$] Scaled equivalent inertia of area $i$
\item[$\Omega_i$] Set of neighbors of area $i$
\end{IEEEdescription}

\noindent\textit{Attack and detection parameters}
\begin{IEEEdescription}[\nomlabelwidth]
\item[$\lambda_r$] Ramp-attack slope
\item[$t_{st},t_{sp}$] Attack start and end times
\item[$\overline{(\cdot)}$] Sample mean for threshold
\item[$\mathcal{D}$] Measurement data buffer
\item[$\lambda_p$] Pulse-attack magnitude
\item[$UB,LB$] Upper and lower detection bounds
\item[$\sigma^2_{(\cdot)}$] Sample variance for threshold
\item[$\mathcal{D}_{th}$] Parameter-history buffer
\item[$g^{(k)}$] CUSUM statistic
\item[$\nu_{kf}, \tau_{kf}$] CUSUM drift \& threshold
\item[$J^{(k)}$] Kalman residual indicator
\item[$\alpha^{(k)}, \alpha$] Dynamic and final ACE scaling factors
\end{IEEEdescription}

\noindent\textit{UIO benchmark symbols}
\begin{IEEEdescription}[\nomlabelwidth]
\item[$\bm{d}(t)$] Lumped unknown input
\item[$E_{sub}$] Unknown-input distribution matrix
\item[$\bm{y}(t)$] UIO measurement vector
\item[$C_{sub}$] UIO subsystem output matrix
\item[$\bm{z}(t)$] UIO internal state vector
\item[$F, L$] UIO matrix and observer gain 
\end{IEEEdescription}

%%%%%%%%%%%%%%%
\section{Introduction}\label{Introduction}

The integration of information technology has significantly improved the operational performance of modern power systems, but it has also increased their vulnerability to cyberattacks, particularly false data injection attacks (FDIAs), which threaten critical functions such as state estimation and automatic generation control (AGC). As a key cyber-physical control function, AGC regulates system frequency and tie-line power deviations using area control error (ACE) signals derived from system measurements. Because these measurements are transmitted through communication networks, they are susceptible to FDIAs, which can distort ACE signals, degrade frequency stability, and potentially trigger cascading failures. To address this issue, a variety of FDIA detection and mitigation strategies have been proposed for AGC systems, including machine-learning-based and model-based approaches.

Regarding machine-learning-based approaches, Zhang \textit{et al.} \cite{Zhang2024} used a Levenberg--Marquardt back-propagation neural network trained on frequency, tie-line power, and active power load data to detect FDIAs. Without using load measurements, He \textit{et al.} \cite{He2020} adopted a supervised classification framework that labels ACE signals as ``normal'' or ``attacked'' to identify FDIAs. In contrast, Musleh \textit{et al.} \cite{Musleh2023} proposed an unsupervised learning approach based on a long short-term memory autoencoder (LSTM-AE), in which reconstruction errors are used to identify FDIAs while reducing human intervention. This unsupervised approach alleviates the limitations of supervised methods, such as the need for extensive labeled data. However, because the LSTM-AE is trained on limited samples, its effectiveness in detecting FDIAs may be restricted, as shown in Section \ref{section: Simulation and Results}.

As for model-based approaches, Khalaf \textit{et al.} \cite{Khalaf2018} and Xiahou \textit{et al.} \cite{Xiahou2021a} applied Kalman-filter-based methods to estimate and mitigate FDIA signals. By verifying the consistency between observed and predicted frequency deviations under observed load changes, Tan \textit{et al.} \cite{Tan2017} identified the sensor data links under attack. Nevertheless, these methods \cite{Khalaf2018,Xiahou2021a,Tan2017} require real-time load observations, which may be impractical and, even when available, increase sensing complexity and data dependency. To relax this requirement, Roy \textit{et al.} \cite{Roy2020} designed a method that leverages forecasted ACE data to identify and mitigate FDIAs. Although this avoids the need for additional load measurement sensors, the method in \cite{Roy2020} did not demonstrate strong resilience to minor ACE signal variations. Ameli \textit{et al.} \cite{Ameli2018} applied an unknown input observer (UIO) to estimate the AGC system states and compute a residual function, where a discrepancy between the residual and a predefined threshold indicates the existence of an FDIA. While the UIO-based detector represents a significant advance by eliminating the need for load observations, it may fail to detect attack models that do not alter the UIO residual, as will be discussed in Section \ref{section: Simulation and Results}.

Beyond the AGC-specific cyberattack literature, likelihood-based parameter estimation for Ornstein-Uhlenbeck (OU) processes is a mature topic in stochastic finance, econometrics, and the statistical inference of diffusion processes \cite{Sorensen2004Survey,Sorensen2009Handbook}. In particular, multivariate OU estimation has been studied in econometrics \cite{Fasen2013MultivariateOU}, while closed-form likelihood expansions and quasi-maximum-likelihood methods have been developed for more general multivariate diffusion models \cite{AitSahalia2008MultivariateDiffusions,Hurn2013QML}. Additional studies have investigated Bayesian and likelihood-based inference for multivariate, graph, and L\'evy-driven OU models \cite{Courgeau2022GraphOU,Lu2022LevyDrivenOU}. More recent work has further examined OU inference under low-frequency observations, measurement noise, and related multivariate continuous-time autoregressive settings \cite{Han2024ModifiedLS,Carter2024MeasurementNoise,Zhang2025LowFrequencyOU,Lucchese2026MCAR}. OU-type and stochastic differential equation models have also been used in power-related applications, such as uncertainty modeling in power systems and wind-power estimation \cite{Verdejo2019PowerSDE,ArenasLopez2020WindPowerOU}. Therefore, the contribution of this paper is not to claim the first derivation of parameters for a multivariate OU process; rather, it lies in developing an application-specific OU-based identification framework for FDIA detection in AGC systems. 

In this study, we propose a maximum likelihood estimation (MLE)-based FDIA detection framework for AGC systems. By modeling the AGC dynamics as a drifted multivariate OU process, we estimate the system state matrix directly from available AGC measurements. Operators can then detect various FDIAs by simply monitoring selected entries of this estimated matrix. The main contributions of this paper are threefold:
\begin{itemize}
\item \textit{Reduced data and modeling dependency}: Unlike existing methods \cite{Tan2017,Khalaf2018,Roy2020,Xiahou2021a}, the proposed detector requires neither real-time load data/forecasts nor detailed AGC system parameters.
\item \textit{Enhanced detection capability}: The proposed method demonstrates superior accuracy, robustness, and speed compared to the UIO \cite{Ameli2018} and LSTM-AE \cite{Musleh2023} detectors. Notably, it successfully identifies coordinated FDIAs that evade UIO detection (see Section \ref{section: Simulation and Results}).
\item \textit{Dynamic thresholding}: A dynamic threshold mechanism is designed to ensure timely FDIA detection while maintaining a low false-alarm rate under fluctuating load conditions.
\end{itemize}

The remainder of this paper is organized as follows. Section II presents the AGC system model. Section III details the proposed FDIA detection algorithm. Section IV evaluates its performance against the UIO and LSTM-AE detectors. Section V concludes the paper. 
%%%%%%%%%%%%%%%%%%%%
\section{The Stochastic Model of the AGC System}\label{section: AGC model}
\subsection{Dynamical Modeling of the AGC System}\label{section:differential function of AGC blocks}
\begin{center}
\begin{figure}[!ht]
\centering
\includegraphics[width=0.45\textwidth]{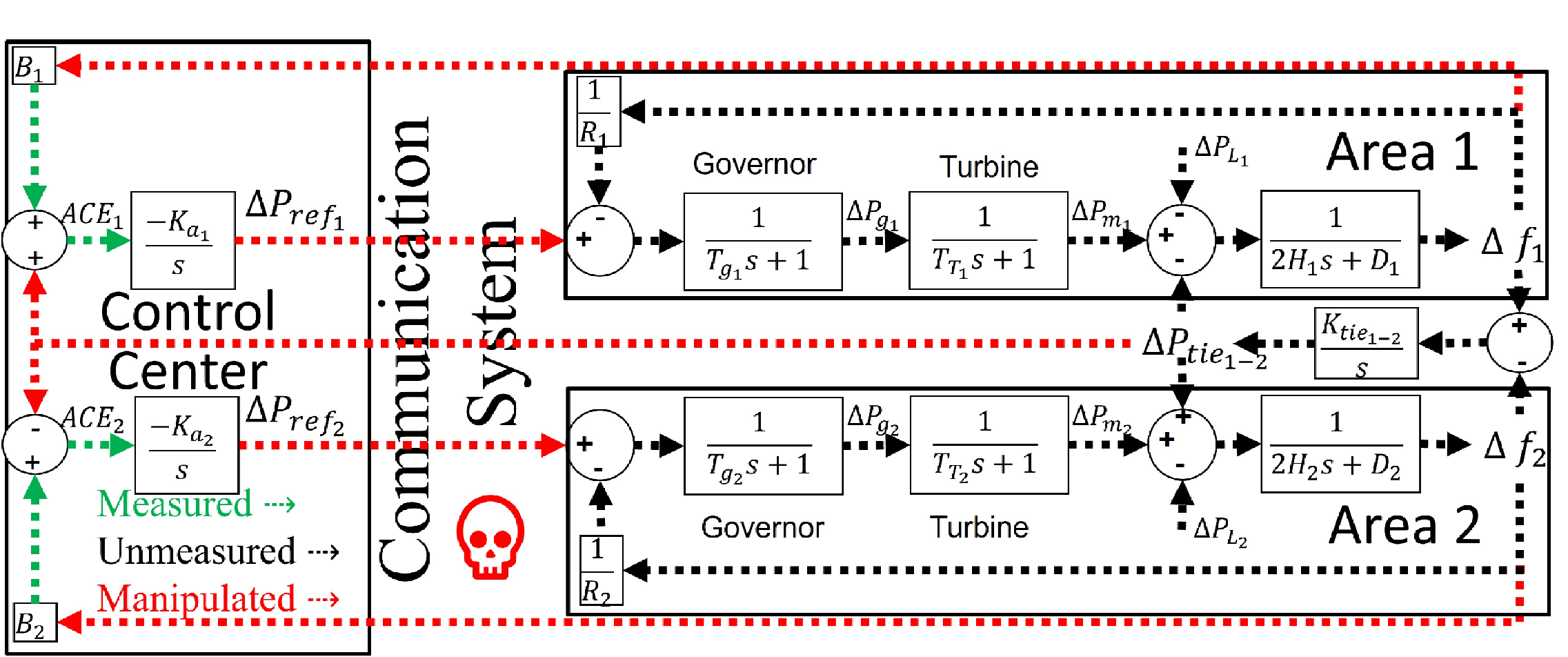}
\caption{The AGC control block of the 2-area system, where green arrows indicate measurable or commanded signals, black arrows represent unmeasured signals, and red arrows denote data that can be manipulated. }
\label{2 areas AGC block detection}
\end{figure}
\end{center}

AGC \cite{Kundur1994} is a vital control system, which monitors the system frequency and adjusts power generation outputs based on sensor measurements to match the power demand and maintain the desired frequency. 
As shown in Fig. \ref{2 areas AGC block detection}, each area of a power system can be represented by equivalent inertia $H_i$, load-damping $D_i$, turbine time constant $T_{T_i}$, governing time constant $T_{g_i}$, and speed regulation $R_i$. The ACE of each area is defined as a linear combination of tie-line error and frequency error with frequency bias $B_i$. The ACE is used to determine the change of power reference point applied to the selected generator units in each area with AGC integrator $K_{a_i}$. Besides, the tie-line power is inherently related to the frequency deviation in interconnected areas \cite{He2020, Khalaf2018}, with tie-line power between Area $i$ and Area $j$ determined by the frequency deviations and coefficient $K_{tie_{i-j}}$. 
Below, we present the dynamic model of the AGC system:
\small
\begin{flalign}
&\Delta {{\dot f}_i} = \frac{1}{{2{H_i}}}(\Delta {P_{{m_i}}} - {D_i}\Delta {f_i}- \sum\limits_{j \in {\Omega _i}} {\Delta {P_{ti{e_{i - j}}}}}  - \Delta {P_{{L_i}}}) \label{eq: fre}\\
&\Delta {{\dot P}_{re{f_i}}} =  -K_{a_i}AC{E_i}   \label{eq: Pref}\\
&\Delta {{\dot P}_{ti{e_{i - j}}}} = {K_{tie_{i - j}}}(\Delta {f_i} - \Delta {f_j}) \label{eq: Ptie}\\
&\Delta {{\dot P}_{{g_i}}} = \frac{1}{{{T_{{g_i}}}}}( - \frac{1}{{{R_i}}}\Delta {f_i} - \Delta {P_{{g_i}}} + \Delta P_{re{f_i}}) \label{eq: Pg}\\
&\Delta {{\dot P}_{{m_i}}} = \frac{1}{{{T_{{T_i}}}}}(\Delta {P_{{g_i}}} - \Delta {P_{{m_i}}}) \label{eq: Pm}\\
&AC{E_i} =  {B_i}\Delta {f_i} + \sum\limits_{j \in {\Omega _i}} {\Delta {P_{ti{e_{i - j}}}}} \label{eq: ACE}
\end{flalign}
\begin{tabular}{ll|ll}
${\Omega_i}$ & connected areas & $H_i$ & equiv. inertia const.\\
$\Delta f_i$ & frequency deviation & $D_i$ & load-damping const.\\
$\Delta P_{g_i}$ & governor power dev. & $R_i$ & speed regulation\\
$\Delta P_{m_i}$ & mech. power dev. & $T_{g_i}$ & governor time const.\\
$\Delta P_{ref_i}$ & power ref. dev. & $T_{T_i}$ & turbine time const.\\
$\Delta P_{tie_{i - j}}$ & tie-line power dev. & $K_{a_i}$ & AGC integrator gain\\
$\Delta P_{L_i}$ & load deviation & $B_i$ & frequency bias factor\\
$K_{tie_{i - j}}$ & tie-line power const. 
\end{tabular}
\normalsize
\subsection{The Stochastic State-Space Model of the AGC System}\label{section:State-space matrices for AGC systems}
In power systems, the OU process is used to model different load phenomena (e.g., \cite{Roberts2016, Mele2016, Du2021}). In this paper, we use the OU mean reversion process (\ref{eq: P_load}) to describe the aggregated power load %at the substation level 
in Area $i$, which converges to and fluctuates around the equilibrium point $\mu_{L_i}$: 
\begin{equation}
\Delta {{\dot P}_{{L_i}}} = -K_{L_i}(\Delta {P_{{L_i}}} - {\mu_{L_i}}) + {\gamma _i}{\xi _i}\label{eq: P_load}
\end{equation}
Particularly,  %where 
${\xi_i}$ is a Gaussian white noise; $\gamma_{i}$ is the diffusion term describing the fluctuation intensity; $K_{L_i}$ is a coefficient describing how quickly $\Delta P_{{L_i}}$ converges to the mean $\mu_{L_i}$. 
For example, in \cite{Mele2016}, $K_{L_i}$ was tuned to $0.02$ per-unit/second (pu/s) to model the stochastic load in \cite{Mele2016}. By analyzing frequency measurements of the All-Island Irish Transmission System, $K_{L_i}$ is set to be $0.0125$ pu/s in \cite{Mele2019}.

Following a load change (e.g., $\mu_{L_i}$ changes from $0$ to another particular $\mu_{L_i}$), the goal of the AGC system is twofold: (i) returning the steady-state frequency error of each area to zero, i.e., $\Delta {f_i}=0$; (ii) ensuring that each area maintains the net tie-line power flow out of the area at its scheduled value, i.e., $\forall j \in {\Omega _i},{\Delta {P_{ti{e_{i - j}}}}}=0$ \cite{Kundur1994}. Therefore, the load change $\mu_{L_i}$ in $\Delta {P_{{L_i}}}$ will result in the same change in $\Delta {P_{{ref_i}}},\Delta {P_{{g_i}}},\Delta {P_{{m_i}}}$ (so that (ii) is satisfied), meaning that the ultimate equilibrium for $[\Delta {f_i},\Delta  {P_{re{f_i}}},\Delta {P_{ti{e_i}}},\Delta {P_{{g_i}}},\Delta {P_{{m_i}}},\Delta {P_{{L_i}}}]$ 
should be $[0,{\mu_{L_i}},0,{\mu_{L_i}},{\mu_{L_i}},{\mu_{L_i}}]$.

Considering these, the dynamic model of the AGC system \eqref{eq: fre}-\eqref{eq: P_load} can be described in the following compact matrix form:

\small
\begin{equation}
\label{eq: AGC_system_compact}
\begin{array}{l}
\underbrace {\left[ {\begin{array}{*{20}{c}}
{\Delta \bm{\dot f}}\\
{\Delta {{\bm{\dot P}}_{ref}}}\\
{\Delta {{\bm{\dot P}}_{tie}}}\\
\hline
{\Delta {{\bm{\dot P}}_g}}\\
{\Delta {{\bm{\dot P}}_m}}\\
{\Delta {{\bm{\dot P}}_L}}
\end{array}} \right]}_{{{\bm{\dot x}}_t}} = A(\underbrace {\left[ {\begin{array}{*{20}{c}}
{\Delta \bm{f}}\\
{\Delta {\bm{P}_{ref}}}\\
{\Delta {\bm{P}_{tie}}}\\
\hline
{\Delta {\bm{P}_g}}\\
{\Delta {\bm{P}_m}}\\
{\Delta {\bm{P}_L}}
\end{array}} \right]}_{{\bm{x}_t}} - \underbrace {\left[ {\begin{array}{*{20}{c}}
0\\
{{\bm{\mu }_L}}\\
0\\
\hline
{{\bm{\mu }_L}}\\
{{\bm{\mu }_L}}\\
{{\bm{\mu }_L}}
\end{array}} \right]}_{{\bm{\mu }_t}}) + \underbrace {\left[ \begin{array}{c}
0 \\
0\\
0\\
0\\
0\\
\Gamma
\end{array} \right]}_S{\bm{\xi }_t}   \\
 A = \\
\left[ {\begin{array}{*{20}{c}}
{\frac{{ - 1}}{2}{H^{ - 1}}D}&0&{\frac{{ - 1}}{2}{H^{ - 1}}V}&\vline& 0&{\frac{1}{2}{H^{ - 1}}}&{\frac{{ - 1}}{2}{H^{ - 1}}}\\
{ - {K_a}B}&0&{ - {K_a}V}&\vline& 0&0&0\\
{{K_t}{V^T}}&0&0&\vline& 0&0&0\\
\hline
{ - {{(R{T_g})}^{ - 1}}}&{T_g^{ - 1}}&0&\vline& { - T_g^{ - 1}}&0&0\\
0&0&0&\vline& {T_T^{ - 1}}&{ - T_T^{ - 1}}&0\\
0&0&0&\vline& 0&0&{-{K_L}}
\end{array}} \right]
\end{array}
\end{equation}
where \small 
\begin{table}[H]
\begin{tabular}{ll}
$\Delta \bm{f} = {[\Delta {f_1},...,\Delta {f_n}]^T}$,&$D = {\text{diag}}[{D_1},...,{D_n}]$,\\
$\Delta {\bm{P}_{ref}} = {[\Delta {P_{re{f_1}}},...,\Delta {P_{re{f_n}}}]^T}$,&$R = {\text{diag}}[{R_1},...,{R_n}]$,\\
$\Delta {\bm{P}_{tie}} = {[\Delta {P_{ti{e_1}}},...,\Delta {P_{ti{e_m}}}]^T}$,&$H = {\text{diag}}[{H_1},...,{H_n}]$,\\
$\Delta {\bm{P}_g} = {[\Delta {P_{{g_1}}},...,\Delta {P_{{g_n}}}]^T}$,&${T_g} = {\text{diag}}[{T _{{g_1}}},...,{T _{{g_n}}}]$,\\
$\Delta {\bm{P}_m} = {[\Delta {P_{{m_1}}},...,\Delta {P_{{m_n}}}]^T}$,&${T_T} = {\text{diag}}[{T _{{T_1}}},...,{T _{{T_n}}}]$,\\
$\Delta {\bm{P}_L} = {[\Delta {P_{{L_1}}},...,\Delta {P_{{L_n}}}]^T}$,&$B = {\text{diag}}[{B_1},...,{B_n}]$,\\
$  \bm{\mu}_L = [{\mu_{L_1},...,\mu_{L_n}}]^T$, & ${K_{a}} = {\text{diag}}[K_{a_1},...,K_{a_n}]$,\\
${\Gamma } = {\text{diag}}[{\gamma _1},...,{\gamma _n}]$, & ${K_{t}} = {\text{diag}}[K_{tie_1},...,K_{tie_m}]$,\\
${K_{L}} = {\text{diag}}[K_{L_1},...,K_{L_n}]$
\end{tabular}
\label{notation}
\end{table}
\noindent \normalsize and $V$ is an ${n \times m}$ oriented incidence matrix, where $n$ and $m$ are the numbers of areas and tie-lines, respectively, such that $ V_{ik}=1,  V_{jk}=-1$ if the $k^{th}$ tie-line transfers power from area $i$ to $j$. Thus, the compact form of the dynamic model can be represented as a drifted multivariate OU mean reversion process \cite{Gardiner2009} that is Gaussian and Markovian as \eqref{eq: OU} 
\begin{equation}
\label{eq: OU}
d{\bm{x}_t} = A({\bm{x}_t} - \bm{\mu}_t)dt + Sd{\bm{W}_t}
\end{equation}
\normalsize where $d{\bm{W}_t} = {\bm{\xi }_t}dt$ and ${\bm{W}_t}$ is the Wiener process with the property that $\mathrm{E}(\bm{W}_t)=0$, ${\bm{W}_t}-{\bm{W}_0} \sim N({0},t)$; the expression of $A$ and $S$ can be found in \eqref{eq: AGC_system_compact}. The system state matrix $A$ carries important information for AGC's operation.
Existing model-based methods \cite{Khalaf2018,Xiahou2021a,Tan2017,Ameli2018} rely on precise knowledge of $A$ such as $K_t$, which may be challenging to obtain accurately.  
Furthermore, it is also challenging to obtain accurate $K_{L_i},\mu_{L_i}$ and thus $\Delta P_{L_i}$ in \eqref{eq: P_load} as discussed in \cite{Hong2016}. 
Therefore, the performance of the methods proposed in \cite{Khalaf2018,Xiahou2021a,Tan2017} may be compromised due to the assumptions of knowing accurate $\Delta P_{L_i}$ values. Although the UIO detector designed in \cite{Ameli2018} relaxes the requirement for $\Delta P_{L_i}$, it may fail to detect certain FDIAs as shown in Section \ref{section: Simulation and Results}.

Given the challenges of existing works, we will propose a novel FDIA detection algorithm in Section \ref{FDIA Algorithm} based on the MLE for the multivariate OU process. Notably, the proposed method can effectively detect various FDIAs without requiring knowledge of model parameters or load predictions.

\section{A Novel FDIA Detection Algorithm}\label{FDIA Algorithm}
We derive the MLE for the OU process to extract AGC parameters from sensor data. When only limited measurements are available, we estimate AGC parameters accordingly. FDIAs disrupt the OU mean reversion, causing deviations in estimated parameters. Using these deviations, we design a novel dynamic FDIA detection algorithm.

\subsection{Estimating the AGC Key Parameters Based on the MLE for the Multivariate OU Process} 
Although the continuous OU mean reversion process is used to model the AGC in \eqref{eq: OU}, power measurements are typically sampled discretely at intervals of $\Delta t$. Assuming ${\bm{\mu }_t}$ is constant and equal to $\bm{\mu }$ within each short period $\Delta t$, the discrete state-space model of \eqref{eq: OU} {can be expressed as:}
\small
\begin{equation}
\label{eq: discrete OU}
{\bm{x}^{(k)}} = {e^{A\Delta t}}{\bm{x}^{(k - 1)}} + (I - {e^{A\Delta t}})\bm{\mu } + \int_0^{\Delta t} {{e^{A(\Delta t - \tau)}}Sd{W_{\tau}}} 
\end{equation}
\normalsize
where ${\bm{x}^{(k)}} \in {\mathbb{R}^{N \times 1}}$ is the $N$ dimensional state vector at time step $k$. Because of the Markovian and Gaussian property of the OU mean reversion process, the conditional distribution function of $\bm{x}^{(k)}$ given ${\bm{x}^{(k-1)}}$ follows the multivariate normal distribution \cite{Meucci2011}:
\small
\begin{equation}
\label{eq: PDF for multi}
{\bm{x}^{(k)}}|{\bm{x}^{(k - 1)}} \sim N({e^{A\Delta t}}{\bm{x}^{(k - 1)}} + \bm{\mu } - {e^{A\Delta t}}\bm{\mu },\Sigma )
\end{equation}
\normalsize
with the probability density function (PDF) as below: 
\small
\begin{align} 
\label{eq: multivariate normal distribution}
\begin{gathered}
  f({\bm{x}^{(k)}}|{\bm{x}^{(k - 1)}};A,\bm{\mu },{\Sigma ^{ - 1}}) = {({(2\pi )^N}\left| \Sigma  \right|)^{ - 1/2}} \\
   %\qquad\qquad\qquad
   \times \exp ( - \frac{1}{2}{({\bm{x}^{(k)}} - {e^{A\Delta t}}({\bm{x}^{(k - 1)}} - \bm{\mu }) - \bm{\mu })^T} \hfill \\
   %\qquad\qquad\qquad
   \times {\Sigma ^{ - 1}}({\bm{x}^{(k)}} - {e^{A\Delta t}}({\bm{x}^{(k - 1)}} - \bm{\mu }) - \bm{\mu })) \hfill \\ 
\end{gathered} 
\end{align}   
\normalsize
where we use the Ito isometry \cite{Mikosch1994} to calculate \small$\Sigma  = \int_0^{\Delta t} {{e^{A(\Delta t - \tau)}}S{S^T}{e^{{A^T}(\Delta t - \tau)}}d\tau}$. \normalsize
%\color{red} The texts span two columns Condensed to one.
Next, the MLE theory of the multivariate OU mean reversion process will be derived and exploited to estimate the parameters $A,\bm{\mu},\Sigma$ purely from the observed $\bm{x}$. Because of the Markovian and Gaussian properties of \eqref{eq: discrete OU}, the log-likelihood function $L(A,\bm{\mu},{\Sigma ^{ - 1}})$ can be defined by taking the logarithm of the product of the PDFs %expressed 
in \eqref{eq: multivariate normal distribution}: 
\small
\begin{align} 
\label{eq: multivariate likelihood}
\begin{gathered}
  L(A,\bm{\mu },{\Sigma ^{ - 1}}) = \log \prod\limits_{k = 1}^M {f({\bm{x}^{(k)}}|{\bm{x}^{(k - 1)}};A,\bm{\mu },{\Sigma ^{ - 1}})}  =  \hfill \\
   - \frac{{MN}}{2}\log 2\pi  - \frac{1}{2}\sum\limits_{k = 1}^M {({{({\bm{x}^{(k)}} - {e^{A\Delta t}}({\bm{x}^{(k - 1)}} - \bm{\mu }) - \bm{\mu })}^T}}  \hfill \\
   \times {\Sigma ^{ - 1}}({\bm{x}^{(k)}} - {e^{A\Delta t}}({\bm{x}^{(k - 1)}} - \bm{\mu }) - \bm{\mu })) - \frac{M}{2}\log \left| \Sigma  \right| \hfill \\ 
\end{gathered} 
\end{align} 
\normalsize
According to the MLE theory, the maximum likelihood estimate of $\hat A, \hat {\bm{\mu}},\hat \Sigma$ should maximize the likelihood function such that the observed data are most likely under the assumed statistical model \cite{Valdivieso2009}:
\begin{equation}
\label{eq: arg_max}
(\hat A,\bm{\hat \mu },{\hat \Sigma ^{ - 1}}) = \arg \max L(A,\bm{\mu },{\Sigma ^{ - 1}};[{\bm{x}^{(1)}},...,{\bm{x}^{(M)}}])
\end{equation}
As a result, we can obtain the maximum likelihood estimates from the observed measurements of $\bm{x}$ if all $\bm{x}$ are accessible, as stated in \textbf{Theorem 1}.

\noindent \textbf{Theorem 1}. (MLE of the multivariate OU mean reversion process)
\label{Thm:globally observable} Consider the multivariate OU mean reversion process in (\ref{eq: OU}), if $M$ measurements of $\bm{x}$, i.e.,  $[{\bm{x}^{(1)}},...,{\bm{x}^{(k)}},...,{\bm{x}^{(M)}}]$, are collected with a sampling time of $\Delta t$, the maximum likelihood estimates for $\hat A, \bm{\hat{\mu}}, \hat \Sigma$ can be obtained by solving a set of simultaneous equations \eqref{eq: derivative of A, mu, sigma 1}-\eqref{eq: derivative of A, mu, sigma 3}: 
\small
\begin{eqnarray} 
\label{eq: derivative of A, mu, sigma 1}
  {e^{\hat A\Delta t}}& = &\sum\limits_{k = 1}^M {(({\bm{x}^{(k)}} - \bm{ \hat \mu }){{({\bm{x}^{(k - 1)}} - \bm{\hat \mu })}^T})}  \hfill \\ \nonumber
   &&\times {(\sum\limits_{k = 1}^M {(({\bm{x}^{(k - 1)}} - \bm{\hat \mu }){{({\bm{x}^{(k - 1)}} - \bm{\hat \mu })}^T})} )^{ - 1}} \hfill \\
\label{eq: derivative of A, mu, sigma 2}
  \bm{\hat \mu }&=&{(I - {e^{\hat A\Delta t}})^{ - 1}}\frac{1}{M}\sum\limits_{k = 1}^M {({\bm{x}^{(k)}} - {e^{ \hat A\Delta t}}{\bm{x}^{(k - 1)}})}  \hfill \\ 
\label{eq: derivative of A, mu, sigma 3}
  \hat \Sigma  &= &\frac{1}{M}\sum\limits_{k = 1}^M {({\bm{x}^{(k)}} - \bm{ \hat \mu } - {e^{\hat A\Delta t}}({\bm{x}^{(k - 1)}} - \bm{\hat \mu }))}  \hfill \\ \nonumber
   &&\times {({\bm{x}^{(k)}} - \bm{\hat \mu } - { e^{\hat A\Delta t}}({\bm{x}^{(k - 1)}} - \bm{ \hat \mu }))^T} \hfill 
\end{eqnarray} 
\normalsize
While likelihood-based inference for multivariate OU models is well-established, Theorem 1 adapts this framework to the AGC setting. The key contribution lies in using this approach to estimate AGC parameters from limited measurements, enabling a robust online FDIA detector.

\subsection{Observability and Vulnerability of Power Measurements and Commands}\label{section:Observability and Vulnerability of Power Measurements and Commands} 
In Fig. \ref{2 areas AGC block detection}, certain internal variables such as governor signal $\Delta \bm{P}_g$ and turbine signal $\Delta \bm{P}_m$ are typically not sent to the control center. 
Conversely, $\Delta \bm{f}$ and $\Delta \bm{P}_{tie}$ are measured and sent to control centers via RTUs (Remote Terminal Units) and PMUs (Phasor Measurement Units), where RTUs use protocols like Modbus, DNP3, and IEC 61850, and PMUs operate with C37.118 or IEC 61850-90-5.
Security weaknesses such as insufficient authentication, encryption, and data integrity make it possible for the intruder to manipulate $\Delta \bm{f}$, $\Delta \bm{P}_{tie}$, and $\Delta \bm{P}_{ref}$.  
In this paper, we assume that $\Delta \bm{P}_g$ and $\Delta \bm{P}_m$ are not measured, while all measurements $\Delta \bm{f},\Delta \bm{P}_{tie},\Delta \bm{P}_{ref}$ can be intercepted and manipulated by the intruder. 
\subsection{Estimating AGC Parameters with Limited Measurements}\label{section:Estimating system parameters with limited measurements in AGC Systems}
Let $\bm{x}_{sub}^{(k)} = {[{(\Delta {\bm{f}^{(k)}})^T},{(\Delta \bm{P}_{ref}^{(k)})^T},{(\Delta \bm{P}_{tie}^{(k)})^T}]^T}$ and define 
\small
\begin{equation}
\label{eq: matrix Q}
Q = \left[ {\begin{array}{*{20}{c}}
  {{I_{n \times n}}}&0&0&\vline & 0&0&0 \\ 
  0&{{I_{n \times n}}}&0&\vline & 0&0&0 \\ 
  0&0&{{I_{m \times m}}}&\vline & 0&0&0 
\end{array}} \right]
\end{equation}
\normalsize
where $I_{n \times n}$ and $I_{m \times m}$ are $n \times n$ and $m \times m$ identity matrices, respectively. Then by \eqref{eq: discrete OU}, we have:
\small
\begin{align}
&\bm{x}_{sub}^{(k)} = Q\bm{x}^{(k)}=\\ \nonumber
&Q({e^{A\Delta t}}{\bm{x}^{(k - 1)}} + (I - {e^{A\Delta t}})\bm{\mu} + \int_0^{\Delta t} {{e^{A(\Delta t - \tau)}}Sd{\bm{W}_\tau})} 
\label{eq: discrete OU x_sub}
\end{align}
\normalsize

\normalsize Since any linear transformation of $\bm{x}$ is also multivariate Gaussian distributed \cite{degroot2012probability}, we can conclude that: 
\small
\begin{equation}
\begin{aligned}
\label{eq: multivariate normal distribution of subsystem}
\bm{x}_{sub}^{(k)}\mid{\bm{x}^{(k - 1)}} \sim N(Q{e^{A\Delta t}}{\bm{x}^{(k - 1)}} + Q\bm{\mu } - Q{e^{A\Delta t}}\bm{\mu },Q\Sigma {Q^T})
\end{aligned}
\end{equation} 
\normalsize
Next, we prove that the term $Q{e^{A\Delta t}}{\bm{x}^{(k - 1)}}$ in (\ref{eq: multivariate normal distribution of subsystem}) can be approximated as ${e^{{A_{sub}}\Delta t}}{\bm{x}_{sub}^{(k - 1)}}$ where
\small
\begin{equation}
\label{eq: approximation1}
\begin{gathered}
  {A_{sub}} = QA{Q^T}, {\bm{\mu }_{sub}} = Q\bm{\mu }\hfill %,{\Sigma _{sub}} = Q\Sigma {Q^T}, \hfill \\
\end{gathered} 
\end{equation}
\normalsize
To see that, the vector $\bm{x_t}$ in \eqref{eq: AGC_system_compact} after discretization and the matrix $A$ partitioned can be expressed as: 
%\color{red} $\bm{x_t}$ in (8), not (k-1) 
%Next, we partition the matrix and vector in \eqref{eq: AGC_system_compact} as
\small
\begin{equation}
\label{eq:Partitioned Matrix}
A = \left[ {\begin{array}{*{20}{c}}
{{A_{sub}}}&\vline& {{A_2}}\\
\hline
{{A_3}}&\vline& {{A_4}}
\end{array}} \right], \bm{x}^{(k - 1)} = \left[ {\begin{array}{*{20}{c}}
{\bm{x}_{sub}^{(k - 1)}}\\
\hline
{\bm{x}_{oth}^{(k - 1)}}
\end{array}} \right]
\end{equation}
\normalsize
If neglecting the contribution of $ \Delta {\bm{P}_m},\Delta {\bm{P}_L}$ on $\Delta \bm{\dot f}$ (see the first row in \eqref{eq: AGC_system_compact}), i.e., setting $A_2=0$, we can obtain:
\small
\begin{equation}
\label{eq:approx_xk-1}
\begin{array}{l}
Q{e^{A\Delta t}}{\bm{x}^{(k - 1)}} \approx Q\exp (\left[ {\begin{array}{*{20}{c}}
{{A_{sub}}}&\vline& 0\\
\hline
{{A_3}}&\vline& {{A_4}}
\end{array}} \right]\Delta t)\left[ {\begin{array}{*{20}{c}}
{\bm{x}_{sub}^{(k - 1)}}\\
\hline
{\bm{x}_{oth}^{(k - 1)}}
\end{array}} \right]\\
 = \left[ {\begin{array}{*{20}{c}}
{{I_{sub}}}&\vline& 0
\end{array}} \right]\left[ {\begin{array}{*{20}{c}}
{{e^{{A_{sub}}\Delta t}}}&\vline& 0\\
\hline
{{A_3}\Delta t + ...}&\vline& {{e^{{A_4}\Delta t}}}
\end{array}} \right]\left[ {\begin{array}{*{20}{c}}
{\bm{x}_{sub}^{(k - 1)}}\\
\hline
{\bm{x}_{oth}^{(k - 1)}}
\end{array}} \right]\\
 = {e^{{A_{sub}}\Delta t}}\bm{x}_{sub}^{(k - 1)}
\end{array}
\end{equation}
\normalsize
Similarly, we can get:
\small
\begin{equation}
Q{e^{A\Delta t}}\bm{\mu } \approx {e^{{A_{sub}}\Delta t}}{\bm{\mu }_{sub}}
\label{eq:approx_mu}
\end{equation}
\normalsize
%\color{red} Is it $\approx$ or $=$?$\approx$ 
Now substituting (\ref{eq: approximation1}), and  (\ref{eq:approx_xk-1})-(\ref{eq:approx_mu}) in (\ref{eq: multivariate normal distribution of subsystem}), we can get an approximated conditional probability distribution:
\small
\begin{equation} 
\label{eq: multivariate normal distribution of subsystem approx}
\begin{aligned}
\bm{x}_{sub}^{(k)}|\bm{x}^{(k - 1)} \sim N({e^{{A_{sub}}\Delta t}}(\bm{x}_{sub}^{(k - 1)} - {\bm{\mu }_{sub}}) + {\bm{\mu }_{sub}},{\Sigma _{sub}})
\end{aligned}
\end{equation} 
\normalsize
where ${\Sigma _{sub}} = Q\Sigma {Q^T}$. (\ref{eq: multivariate normal distribution of subsystem approx}) will degrade to \eqref{eq: multivariate normal distribution of subsystem approx_sub} %=P(\bm{x}_{sub}^{(k)}|\bm{x}_{sub}^{(k - 1)})$: 
since the mean and variance depend only on $\bm{x}_{sub}$: 
\small
\begin{equation} 
\label{eq: multivariate normal distribution of subsystem approx_sub}
\begin{aligned}
\bm{x}_{sub}^{(k)}|\bm{x}_{sub}^{(k - 1)} \sim N({e^{{A_{sub}}\Delta t}}(\bm{x}_{sub}^{(k - 1)} - {\bm{\mu }_{sub}}) + {\bm{\mu }_{sub}},{\Sigma _{sub}})
\end{aligned}
\end{equation} 
\normalsize
Therefore, similar to the proof of Theorem 1 based on (\ref{eq: PDF for multi}), we can estimate the parameters $A_{sub}$, $\bm{\mu}_{sub}$, $\Sigma_{sub}$
from the obtained $\bm{x}_{sub}$. Specifically, replacing $\bm{x},A,\bm{\mu },\Sigma $ by ${\bm{x}_{sub}},{A_{sub}},{\bm{\mu }_{sub}},{\Sigma _{sub}}$ in \eqref{eq: derivative of A, mu, sigma 1}-\eqref{eq: derivative of A, mu, sigma 3}, %\textbf{Theorem 1}, 
power system operators can obtain good estimates of ${{\hat A}_{sub}},{{\bm{\hat \mu }}_{sub}}, {{\hat \Sigma }_{sub}}$.

It should be noted that although some approximations are made in the above derivations, the simulation results in Section \ref{section: Simulation and Results} demonstrate that the accuracy of $\hat{A}_{sub}$ remains reasonably high using only $\bm{x}_{sub}$ when the power system is not compromised through intrusions and/or experiences natural load variations. Once the power system is under cyberattacks, the estimated parameters in $\hat A_{sub}$ will deviate from the pre-attack model parameters, as the AGC dynamics no longer follow \eqref{eq: multivariate normal distribution of subsystem approx_sub}. 
Leveraging this, the subsequent subsection will propose a novel FDIA detection algorithm that effectively and promptly identifies FDIAs by monitoring the estimated parameters in $\hat{A}_{sub}$. 
\subsection{The Proposed Online FDIA Detection Algorithm}\label{section:Different FDIA Types and Unified Detection Approach}
Before presenting the detailed detection algorithm, we introduce the basic rules that power system operators may adopt to detect intrusions and cyberattacks. Some typical FDIAs will also be introduced. In the control room, an alarm might be raised if the calculated ACE exceeds 0.1 per unit (pu) \cite{Sridhar2014}. Therefore, we assume that the intruder is smart enough to launch stealth cyberattacks that obey the basic rule:
\begin{equation}
\label{eq: basic rules}
\forall i \in [1,...,n],  -0.1\text{ pu}<ACE_i^{(k)} < 0.1\text{ pu}
\end{equation}

In addition, different templates of FDIAs (e.g., ramp attacks, pulse attacks) on AGC were discussed in previous works \cite{Sridhar2014, He2020}. 
Ramp attacks gradually alter true measurements by adding a ramp function that increases or decreases over time. 
\begin{equation}
\label{eq: ramp_FDIA_abstract}
{\tilde{\bm{x}}_t} = \left\{ {\begin{array}{*{20}{c}}
{0,{\text{ for }}t < t_{st}{\text{ or }}t > t_{sp}}\\
{\;{\lambda _r}(t - {\text{ }}t_{st}),{\text{ for }}t_{st} \le t \le t_{sp}}
\end{array}} \right.
\end{equation}
where $t$, $t_{st}$, and $t_{sp}$ represent the time, the initiation time of the FDIA, and the end time of the FDIA, respectively. Assuming discrete injection of FDIA with the interval $\Delta t$, we can transform $\tilde{\bm{x}}_t$ to $\bm{\tilde {\bm{x}}}^{(k)}$ using the relation $k = \left\lfloor {{\text{ }}t/\Delta t} \right\rfloor$, where $ \left\lfloor\right\rfloor$ is the floor function. Pulse attacks involve abrupt modification of true measurements by adding a fixed value as a fixed pulse with parameter $\lambda _p$.

\begin{equation}
\label{eq: pulse_FDIA_abstract}
{\bm{\tilde {\bm{x}}}_t} = \left\{ {\begin{array}{*{20}{c}}
{0,{\text{ for }}t < t_{st}{\text{ or }}t > t_{sp}}\\
{\;{\lambda _p},{\text{ for }}t_{st} \le t \le t_{sp}}
\end{array}} \right.
\end{equation}
Similarly, assuming pulse attacks are injected discretely with the interval $\Delta t$, we can convert $\tilde{\bm{x}}_t$ into $\bm{\tilde {\bm{x}}}^{(k)}$ using the same relation $k = \left\lfloor {{\text{ }}t/\Delta t} \right\rfloor$.

When considering the effect of these attacks on different states of the AGC model, it is clear that the model's behavior before and post-FDIA can vary considerably. FDIAs targeting $\Delta \bm{f}$ or $\Delta \bm{P}_{tie}$ manipulate the reported measurements, not the actual measurements. Hence, the model changes from \eqref{eq: OU} to \eqref{eq: false data injection attack_freorPtie_SDE}, indicating that only the command $\Delta \bm{P}_{ref}$ will be affected by the FDIA but not the true measurements. 
\begin{equation}
\label{eq: false data injection attack_freorPtie_SDE}
d\bm{x}_t = A(\bm{x}_t - \bm{\mu}_t)dt + A^{P_{ref}}\tilde{\bm{x}}_t dt + S d\bm{W}_t
\end{equation}
Particularly, 
\small
\footnotesize
\begin{equation}
\label{eq: false data injection attack_freorPtie auxiliary}
\begin{array}{l}
{A^{{P_{ref}}}} = \left[ {\begin{array}{*{20}{c}}
{A_{sub}^{{P_{ref}}}}&\vline& 0\\
\hline
0&\vline& 0
\end{array}} \right],\mbox{where }A_{sub}^{{P_{ref}}} = \left[ {\begin{array}{*{20}{c}}
0&0&0\\
{ - {K_a}B}&0&{ - {K_a}V}\\
0&0&0
\end{array}} \right]\\
{{\bm{\tilde {\bm{x}}}}^{(k - 1)}} = {\left[ {\bm{\tilde {\bm{x}}}_{sub}^{(k - 1)},\bm{0}} \right]^T},\mbox{where }\bm{\tilde {\bm{x}}}_{sub}^{(k - 1)} = {\left[ {\Delta {{\bm{\tilde f}}^{(k - 1)}},\bm{0},\Delta \bm{\tilde P}_{tie}^{(k - 1)}} \right]^T} \nonumber
\end{array}
\end{equation} 
\normalsize
and $\bm{\tilde {\bm{x}}}^{(k - 1)}$ represents the injected false measurements that may follow a specific attack template discussed above. Then the sampled discrete model of \eqref{eq: false data injection attack_freorPtie_SDE} is:
\small
\begin{flalign}
\label{eq: false data injection attack_freorPtie}
{\bm{x}^{(k)}} &= {e^{A\Delta t}}{\bm{x}^{(k - 1)}} + (I - {e^{A\Delta t}})\bm{\mu } + \int_0^{\Delta t} {{e^{A(\Delta t - \tau)}}Sd{W_{\tau}}} \\
 &+ \int_0^{\Delta t} {{e^{A(\Delta t - \tau)}}{A^{{P_{ref}}}}} {{\bm{\tilde {\bm{x}}}}^{(k - 1)}}d\tau \nonumber
\end{flalign}
\normalsize
Based on \eqref{eq: false data injection attack_freorPtie}, an approximate conditional PDF similar to \eqref{eq: multivariate normal distribution of subsystem approx_sub} can be obtained during FDIAs targeting $\Delta \bm{f}$ or $\Delta \bm{P}_{tie}$:
\small
\begin{flalign}
\label{eq: PDF for multi after FDIA}
\bm{x}_{sub}^{(k)} \mid \bm{x}_{sub}^{(k - 1)} \sim &N({e^{{A_{sub}}\Delta t}}(\bm{x}_{sub}^{(k - 1)} - {\bm{\mu }_{sub}}) + {\bm{\mu }_{sub}} + \\ \nonumber
&\int_0^{\Delta t} {{e^{{A_{sub}}(\Delta t - \tau)}}A_{sub}^{{P_{ref}}}} \bm{\tilde {\bm{x}}}_{sub}^{(k - 1)}d\tau,{\Sigma _{sub}}) 
\end{flalign}
\normalsize
On the other hand, FDIAs targeting $\Delta \bm{P}_{ref}$ lead to altered command $\Delta \bm{P}_{ref}$ without operators' awareness, which will affect both the system's true measurements and command. 
\small
\begin{flalign}
\label{eq: false data injection attack_Pref_SDE}
d{\bm{x}_t} = A({\bm{x}_t} + {\bm{\tilde {\bm{x}}}_t} - {\bm{\mu }_t})dt + Sd{\bm{W}_t}
\end{flalign}
\begin{flalign}
\label{eq: false data injection attack_Pref}
{\bm{x}^{(k)}} &= e^{A\Delta t}({\bm{x}^{(k - 1)}} + {\bm{\tilde {\bm{x}}}^{(k - 1)}}) + (I - {e^{A\Delta t}})\bm{\mu } \\ 
&+ \int_0^{\Delta t} {{e^{A(\Delta t - \tau)}}Sd{W_\tau}} \nonumber
\end{flalign}
\normalsize
Likewise, we can derive an approximate conditional PDF: 
\begin{flalign}
\label{eq: PDF for multi after FDIA_Pref}
\bm{x}_{sub}^{(k)} \mid \bm{x}_{sub}^{(k - 1)} \sim &N({e^{{A_{sub}}\Delta t}}(\bm{x}_{sub}^{(k - 1)} - {\bm{\mu }_{sub}}) + {\bm{\mu }_{sub}} + \\ \nonumber
&{e^{{A_{sub}}\Delta t}}\bm{\tilde {\bm{x}}}_{sub}^{(k - 1)},{\Sigma _{sub}})
\end{flalign}
\normalsize
The deviations in estimated system parameters post-FDIA, as observed in equations (\ref{eq: PDF for multi after FDIA}) and (\ref{eq: PDF for multi after FDIA_Pref}), enable the development of a dynamic FDIA detection algorithm. A dynamic threshold mechanism is used to minimize false alarms while preserving sensitivity to true anomalies.

\begin{algorithm}
\caption{Online FDIA Detection based on the MLE of multivariate OU mean reversion process}
\begin{algorithmic}
\Inputs{MLE window $M=300$; threshold window $M_{th}=3000$; Empty databases $\mathcal{D}$ and $\mathcal{D}_{th}$; index $k \leftarrow 0$; ${\text{flag}} \leftarrow 1$. } 
\Outputs{Detection or Non-detection of FDIA}
\Function{UpdateThresholds}{$\mathcal{D}_{th}$}
    \State Extract  ${K_{{a_i}}}{B_i}$, ${K_{{a_i}}}$, ${K_{{t_i}}}$ from the latest $M_{th}$ matrices $\hat A_{sub}$ 
    in $\mathcal{D}_{th}$, %i.e., $[\hat A_{sub}^{(k-M_{th}+1)},...,\hat A_{sub}^{(k)}]$ in $\mathcal{D}_{th}$, 
    calculate their mean ($\overline{{\cdot}}$) and variance ($\sigma^2_{\cdot}$) for ${K_{{a_i}}}{B_i}$, ${K_{{a_i}}}$, ${K_{{t_i}}}$  across all $n$ areas, i.e., $\forall i \in [1,...,n]$, 
    \State Update thresholds $({K_{{a_i}}}{B_i})^{UB,LB} = \overline{{K_{{a_i}}}{B_i}}  \pm 4\sigma_{{K_{{a_i}}}{B_i}}$, $K_{a_i}^{UB,LB} = \overline{K_{a_i}} \pm 4\sigma_{K_{a_i}}$, $K_{tie_i}^{UB, LB} = \overline{K_{tie_i}} \pm 4\sigma_{K_{tie_i}}$
\EndFunction
\While{$k < M + M_{th}$ } \Comment{Initialization stage}
    \State $k \leftarrow k+1$ 
    \State Add $\bm{x}_{sub}^{(k)} = {[{(\Delta {\bm{f}^{(k)}})^T},{(\Delta \bm{P}_{ref}^{(k)})^T},{(\Delta \bm{P}_{tie}^{(k)})^T}]^T}$ to $\mathcal{D}$
    \If{$k \geq M$}
        \State Based on $[{\bm{x}_{sub}^{(k - M+1)}},...,{\bm{x}_{sub}^{(k)}}]$ in $\mathcal{D}$, estimate $\hat A_{sub}^{(k)}$ using \eqref{eq: derivative of A, mu, sigma 1}-\eqref{eq: derivative of A, mu, sigma 3} in \textbf{Theorem 1}; add $\hat A_{sub}^{(k)}$ to $\mathcal{D}_{th}$
    \EndIf
\EndWhile
\While{$k \geq M + M_{th}$ and flag} \Comment{ Detection stage}
    \State $k \leftarrow k+1$ 
    \State Add $\bm{x}_{sub}^{(k)} = {[{(\Delta {\bm{f}^{(k)}})^T},{(\Delta \bm{P}_{ref}^{(k)})^T},{(\Delta \bm{P}_{tie}^{(k)})^T}]^T}$ to $\mathcal{D}$
     \State Based on $[{\bm{x}_{sub}^{(k - M+1)}},...,{\bm{x}_{sub}^{(k)}}]$ in $\mathcal{D}$, estimate $\hat A_{sub}^{(k)}$ using \eqref{eq: derivative of A, mu, sigma 1}-\eqref{eq: derivative of A, mu, sigma 3} in \textbf{Theorem 1}; add $\hat A_{sub}^{(k)}$ to $\mathcal{D}_{th}$
    \State \Call{UpdateThresholds}{$\mathcal{D}_{th}$}
    \If {$\exists i \in [1,...,n]$, at least one of ${({K_{{a_i}}}{B_i})^{(k)}}$, ${K_{{a_i}}^{(k)}}$, ${K_{{t_i}}^{(k)}}$ exceed respective thresholds}
    \State $\text{flag} \leftarrow  0$; If no topological fault occurs, raise an alarm and all generators will work in local mode
    \EndIf
\EndWhile
\end{algorithmic}
\end{algorithm}
%%%%%%%%%%%%%%
\section{Simulations}\label{section: Simulation and Results}
In simulation, we will assess the efficacy of the proposed MLE detection algorithm using a 2-area benchmark system\cite{Khalaf2018,Saadat2015}. 
Specifically, various ramp FDIAs and pulse FDIAs with different attack intensities and targeting different parameters are considered. 
The performance of the proposed MLE detector is also compared with two existing detectors, namely, the model-based UIO detector from \cite{Ameli2018} and the data-driven LSTM-AE detector from \cite{Musleh2023}. The simple and sophisticated FDIAs are designed to subtly modify the measurement while adhering to the basic rules (\ref{eq: basic rules}), making detection challenging. 
%%%%%%%%%%%%%%%%%%%%%%%%%%%%%%%%%%%%%%%%%%%%%
\subsection{Benchmark Systems and Configuration of MLE, UIO, and LSTM-AE Detectors}\label{section: Benchmark}

The 2-area system is presented in Fig. \ref{2 areas AGC block detection}, with the AGC parameters given in Table \ref{table: Parameters of the 2-area system} (obtained from Fig. 12.26 in \cite{Saadat2015}). Note that the values selected for $\Gamma$ ensure relatively small Area Control Error (ACE) and tie-line power variations under normal operating conditions, obeying the basic rules in \eqref{eq: basic rules}. 

\begin{table}[H]
\centering
\caption{\textsc{2-Area Benchmark}  (Fig. 12.26 in \cite{Saadat2015} )}
\label{table: Parameters of the 2-area system}
\resizebox{\linewidth}{!}
{
\begin{tabular}{|c|c|c|c|c|c|c|c|c|}
\hline
 & $H_i$ (s) & $D_i$ & $R_i$ (pu) & $T_{g_i}$ (s) & $T_{T_i}$ (s)& $K_{a_i}$ & $B_i$ & $K_{L_i}$\\
\hline
Area 1 & 5 & 0.6 & 0.05 & 0.2 & 0.5 & 0.3 & 20.6 & 0.005\\
\hline
Area 2 & 4 & 0.9 & 0.0625 & 0.3 & 0.6 & 0.3 & 16.9  & 0.005\\
\hline
Tie-line &\multicolumn{4}{c}{$K_{tie_{1-2}}=2$} & \multicolumn{4}{|c|}{Base Power = 1000 MVA}\\
\hline
\end{tabular}
}
\end{table}

For the MLE detector, the accessible states and subsystem parameters are defined in \eqref{eq: accessible states in 2-area system}--\eqref{eq: Drift subvector of 2 area AGC}. 
\small
\begin{align}
\label{eq: accessible states in 2-area system}
\bm{x}_{sub}^{(k)} &= {[\Delta f_1^{(k)},\Delta f_2^{(k)},\Delta P_{re{f_1}}^{(k)},\Delta P_{re{f_2}}^{(k)},\Delta P_{ti{e_{1 - 2}}}^{(k)}]^T} \\
\label{eq: Submatrix A of 2 area AGC}
A_{sub} &= \left[ {\begin{array}{*{20}{c}}
{- {D_1}/2{H_1}}&0&0&0&{{ - 1}/{2{H_1}}}\\
0&{- {D_2}/2{H_2}}&0&0&{{ 1}/{2{H_2}}}\\
{ - {K_{a_1}}{B_1}}&0&0&0&{ - {K_{a_1}}}\\
0&{ - {K_{a_2}}{B_2}}&0&0&{{K_{a_2}}}\\
{{K_{tie_{1 - 2}}}}&{ - {K_{tie_{1 - 2}}}}&0&0&0
\end{array}} \right]\\
\label{eq: Drift subvector of 2 area AGC}
{\bm{\mu }_{sub}} &= {\left[ {\begin{array}{*{20}{c}}
  0&0&{{\mu _{{L_1}}}}&{{\mu _{{L_2}}}}&0 
\end{array}} \right]^T}
\end{align}
\normalsize
%%%%%%%%%%%%%%%%%

For comparison purposes, the UIO detector is implemented on a locally observable 5-state subsystem extracted from the 9-state linearized AGC model used in MATLAB. The retained states correspond to the continuous-time counterparts of \eqref{eq: accessible states in 2-area system}, and the associated subsystem matrix is \(A_{sub}\) in \eqref{eq: Submatrix A of 2 area AGC}. By lumping the omitted mechanical-power states into a net unknown input, \(\bm{d}(t) \triangleq \Delta \bm{P}_L(t) - \Delta \bm{P}_m(t)\), the continuous-time subsystem and measurement model are written as
\begin{align}
\label{eq: UIO system_sub_cont}
\dot{\bm{x}}_{sub}(t) &= A_{sub}\bm{x}_{sub}(t) + E_{sub}\bm{d}(t)\\
\label{eq: UIO measurement_cont}
\bm{y}(t) &= C_{sub}\bm{x}_{sub}(t)
\end{align}
with \(C_{sub}=I_{5}\) and \(E_{sub} = \left[ \begin{smallmatrix} -\frac{1}{2H_1} & 0 & 0 & 0 & 0 \\ 0 & -\frac{1}{2H_2} & 0 & 0 & 0 \end{smallmatrix} \right]^T\). Therefore, the UIO existence condition is satisfied as \(\text{rank}(C_{sub}E_{sub}) = \text{rank}(E_{sub}) = 2\).
The standard continuous-time UIO \cite{Ameli2018} implemented in MATLAB is
\begin{align}
\label{eq: UIO observer_z}
\dot{\bm{z}}(t) &= F\bm{z}(t) + L\bm{y}(t)\\
\label{eq: UIO observer_x}
\hat{\bm{x}}_{sub}(t) &= \bm{z}(t) + E_{sub}(C_{sub}E_{sub})^+\bm{y}(t)
\end{align}
where the observer gain $L$ is obtained by pole placement with desired poles $[-10,-20,-30,-40,-50]$, such that $F = (I - E_{sub}(C_{sub}E_{sub})^+C_{sub})A_{sub} - LC_{sub}$.
%%%%%%%%%%%%%%%%%%%%%%%%%%%%%

The LSTM-AE detector \cite{Musleh2023} is implemented using MATLAB's \textit{deepSignalAnomalyDetector} function. The encoder incorporates a 6-unit LSTM layer followed by a 3-unit LSTM layer, while the decoder features a 3-unit LSTM layer followed by a 6-unit LSTM layer. The measurements \small$[{(\Delta {\bm{f}^{(k)}})^T}, {(\Delta \bm{P}_{ref}^{(k)})^T}, {(\Delta \bm{P}_{tie}^{(k)})^T}]^T$\normalsize are z-score normalized. Subsequently, the LSTM-AE is trained over 1000 epochs with a learning rate of 0.001 using the Adam optimizer.

%%%%%%%%%%%%%%%%%%%%%%%%%%%%%%%%%%%%%%%%%%%%%
\subsection{Threshold Selection and Load Fluctuations on FPR of MLE, UIO, and LSTM-AE Detectors in No-Attack Scenarios}\label{sec:threshold selection}
All detection methods exhibit trade-offs between FP rate and detection time across different thresholds. Stricter thresholds reduce false positives at the cost of increased detection time, while looser thresholds enable faster attack detection but with more false alarms. For fair comparison, we calibrated each detector's threshold to achieve approximately 1\% FP rate under attack-free conditions with baseline load parameters ${\bm{\mu}_L} = {[0,0]^T}$ and $\bm{\gamma} = [0.005; 0.005]$.
The selected thresholds were: $4\sigma$ dynamic threshold for MLE, $3.5\sigma$ dynamic threshold for UIO, and 99th percentile threshold for LSTM-AE.
Power systems naturally experience load variations, which may affect the performance (particularly the false positive rate) of detectors. To investigate this, we vary the long-term and short-term parameters of the dynamic loads, i.e., $\mu_{L_i}$ and $\gamma_i$ in (\ref{eq: P_load}), mimicking slow and fast changes of loads.

\begin{table}[htbp]
\centering
\caption{False Positive Rate (\%) Under Different Load Conditions Without Cyberattack}
\label{tab:fp_rates_load_fluctuations}
\begin{tabular}{ccccc}
\hline
\multicolumn{2}{c}{\textbf{Load Conditions}} & \multicolumn{3}{c}{\textbf{False Positive Rate (\%)}} \\
\cline{1-2} \cline{3-5}
$\mu_L$ & $\gamma$ & MLE & UIO & LSTM-AE \\
\hline
$[0; 0]$ & $[0.005; 0.005]$ & 1.1 & 1.0 & 1.0 \\
$[0.1; 0]$ & $[0.005; 0.005]$ & 1.0 & 1.0 & 7.1 \\
$[0.1; 0.1]$ & $[0.005; 0.005]$ & 0.8 & 1.0 & 2.8 \\
$[0; 0]$ & $[0.01; 0.005]$ & 0.7 & 1.5 & 42.8 \\
$[0; 0]$ & $[0.01; 0.01]$ & 1.2 & 3.7 & 59.4 \\
\hline
\end{tabular}
\end{table}

Row~1 in Table~\ref{tab:fp_rates_load_fluctuations} provides the baseline ($\bm{\mu}_L=[0;0]$, $\bm{\gamma}=[0.005;0.005]$) for comparing detector performance under different parameter variations. Rows~2--3 test long-term parameter $\bm{\mu}_L$ changes while Rows~4--5 evaluate short-term parameter $\bm{\gamma}$ variations. The proposed MLE detector demonstrates good stability, maintaining a consistent 1\% FP rate across all conditions. In contrast, the UIO detector shows FP rate degradation to 3.7\% when the short-term parameter $\bm{\gamma}$ changes to $[0.01;0.01]$, while the LSTM-AE exhibits severe sensitivity, with FP rates even reaching 59.4\% (Row~5). These results show the superior stability of model-based approaches (MLE and UIO) compared to the purely data-driven LSTM-AE method, which presents greater susceptibility to load variations.
%%%%%%%%%%%%%%%%%%%%%%%%%%%%%%%%%%%%%%%%%%
\subsection{Detect Single Ramp FDIAs in 2-Area Systems} \label{Detect Single Ramp/Pulse FDIAs}
\begin{figure}[htbp]
\centering
\includegraphics[width=0.45\textwidth]{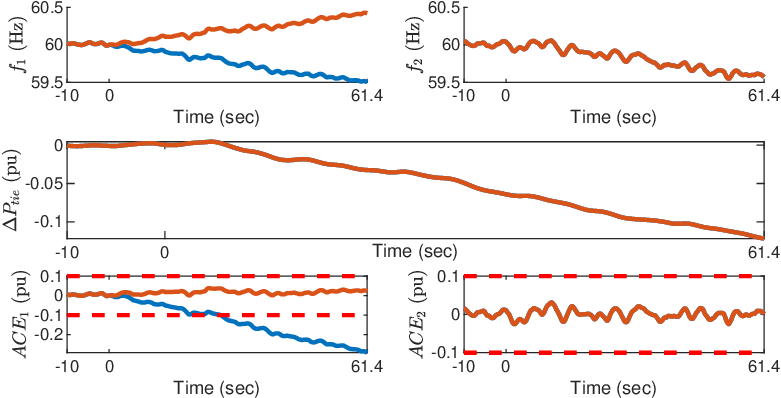}
\caption{$f$, $\Delta P_{tie}$, and $ACE$ before and during the ramp FDIA with $\lambda_r=5\times 10^{-5}$ for $\Delta f_1$. Blue: True trajectories, where undetected FDIA drives $f_1$ to 59.5 Hz; Red: Manipulated trajectories in the control center; Red dashed: ACE alarm threshold. Red trajectories stay within thresholds, avoiding alarm.} 
\label{Fre_Ptie pre and post ramp FDIA}
\end{figure}
We first consider ramp FDIAs. 
\small
\begin{align}
\label{eq: Single Ramp attack}
\begin{array}{l}
{{\bm{\tilde x}}^{(k - 1)}} = {[\Delta \tilde f_1^{(k - 1)},\bm{0}_{1 \times 10}]^T}\\
\Delta \tilde f_1^{(k - 1)} = \left\{ {\begin{array}{*{20}{c}}
{0,{\text{ for }}t < {t_{st}}{\text{ or }}t > {t_{sp}}}\\
{{\lambda _r}(t - {t_{st}}),{\text{ for }}{t_{st}} \le t \le {t_{sp}}}
\end{array}} \right.\\
{\text{where }}{\lambda _r} = 5\times 10^{-5},{t_{st}} = 0{\text{ s}},{t_{sp}} = 600{\text{ s}}
\end{array}
\end{align}
\normalsize
Assuming the intruder injects the ramp attack into $\Delta {{\tilde f}_1}$ as \eqref{eq: Single Ramp attack}. The ramp FDIA is initiated at $t_{st} = 0 \text{ s}$ and stops at $t_{sp} = 600 \text{ s}$. The value for $\lambda_r$ is intentionally chosen to be small enough to meet the basic rules (\ref{eq: basic rules}) set by the bad data detection (BDD),  ensuring the FDIA remains stealthy and challenging to detect. %
As shown in Fig. \ref{Fre_Ptie pre and post ramp FDIA}, if the FDIA goes undetected, the true frequency in Area 1 will drop below the critical threshold of 59.5 Hz at 61.4 s. 
Operating the system at such low frequencies can cause significant damage to synchronous equipment, trigger load shedding, and even result in widespread power outages. However, due to the small attack amplitude, the operator may not notice even when the true frequency, the blue trajectory in Fig.~\ref{Fre_Ptie pre and post ramp FDIA}, reaches 59.5 Hz, which may trigger load shedding or local protection mechanisms. %assuming 
Note that the basic detection rules (\ref{eq: basic rules}) will not detect this attack as the manipulated ACE values (the red trajectories in Fig.~\ref{Fre_Ptie pre and post ramp FDIA}) are always within the limits.

\begin{figure}[htbp]
\centering
\begin{subfigure}{0.16\textwidth}
  \centering
  \includegraphics[width=\textwidth]{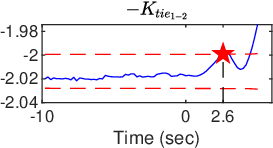}
  \caption{MLE}
  \label{MLE_FDIA_ramp_fre1}
\end{subfigure}%
\begin{subfigure}{0.16\textwidth} 
  \centering
  \includegraphics[width=\textwidth]{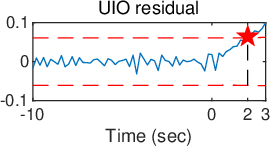}
  \caption{UIO \cite{Ameli2018}}
  \label{UIO_residual_single_ramp_FDIA_fre}
\end{subfigure}%
\begin{subfigure}{0.16\textwidth} 
  \centering
  \includegraphics[width=\textwidth]{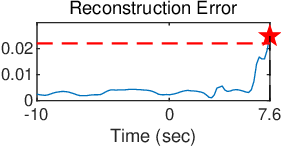}
  \caption{LSTM-AE \cite{Musleh2023}}
  \label{ramp_fre1_AI}
\end{subfigure}
\caption{Comparative responses (detection indicators) to the ramp FDIA with $\lambda_r=5\times 10^{-5}$ for $\Delta f_1$. Red dashed lines: thresholds; red star: FDIA is detected.}
\label{Tie-line FDIA for UIO and LSTM-AE}
\end{figure}
%%%%%%%%%%%%%%%%%%%%%%%%%%%%%%%%%%%%
Fig.~\ref{MLE_FDIA_ramp_fre1} shows that the proposed MLE detector (\textbf{Algorithm 1}) detects the ramp FDIA in 2.6 s, triggered when $-K_{tie_{1-2}}$ in $\hat{A}_{sub}$ exceeds its bounds. For comparison, the UIO (Fig.~\ref{UIO_residual_single_ramp_FDIA_fre}) and LSTM-AE (Fig.~\ref{ramp_fre1_AI}) detectors identify the same attack in 2.0 s and 7.6 s, respectively. 

\begin{figure}[H]
    \centering
    \includegraphics[width=0.45\textwidth]{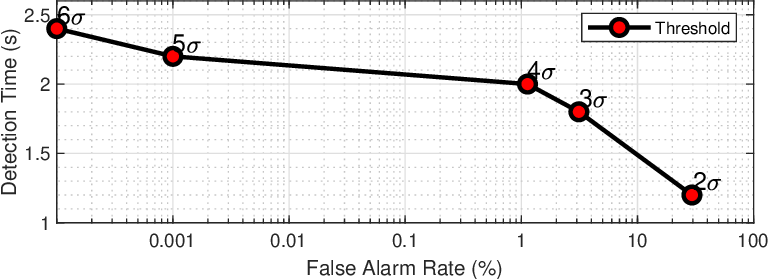}
    \caption{Detection time versus FP rate (log scale) under the ramp FDIA with $\lambda_r=5\times 10^{-5}$ for $\Delta f_2$.}
\label{fig:DetectionTime_vs_FalseRate_log reviewer 1}
\end{figure}

As discussed in Section \ref{sec:threshold selection}, detectors trade off FP rate and detection time across thresholds. Fig. \ref{fig:DetectionTime_vs_FalseRate_log reviewer 1} shows these for ramp FDIA ($\lambda_r=5 \times 10^{-5}$) on $\Delta f_2$. The 4$\sigma$ threshold balances reasonable detection time and low FP rate.

%%%%%%%%%%%%%%%%%%%%%%%%%%%%%%%%%%%%%%%%%%
%%%%%%%%%%%%%%%%%%%%%%%%%%%%%%%%%%%%%%%%%%
\subsection{Detect Sophisticated Attacks in 2-Area Systems} \label{Detection and Analysis of Coordinated Ramp_Pulse FDIAs}
\begin{figure}[htbp]
\centering
\includegraphics[width=0.5\textwidth,keepaspectratio=true]{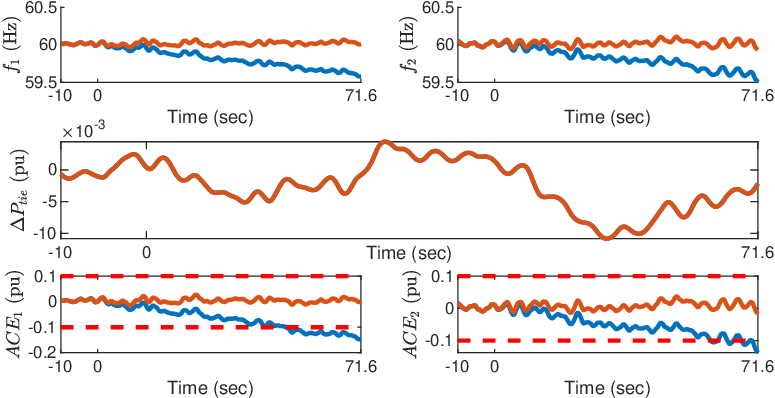}
\caption{$f$, $\Delta P_{tie}$, and $ACE$ before and during the %undetected  
coordinated FDIA with $\lambda_r = 2\times 10^{-5}$ for both $\Delta f_1$ and $\Delta f_2$. Blue: True trajectories, where undetected FDIA drives $  f_1  $ to 59.5 Hz; Red: Manipulated trajectories in control center; Red dashed: ACE alarm threshold. Red trajectories stay within thresholds, avoiding alarm.}
\label{Fre_Ptie pre and post Coordinated Ramp Attacks}
\end{figure}

This subsection examines coordinated ramp FDIAs targeting two measurements simultaneously in the 2-area system, which was proposed in \cite{He2020}. The attack aims to manipulate both $\Delta {f_1}$ and $\Delta {f_2}$ as described in \eqref{eq: coordinate ramp attack}:
\small
\begin{equation}
\label{eq: coordinate ramp attack}
\begin{array}{l}
{{\bm{\tilde x}}^{(k - 1)}} = {[\Delta \tilde f_1^{(k - 1)},\Delta \tilde f_2^{(k - 1)},{\bm{0}_{1 \times 9}}]^T}\\
\Delta \tilde f_1^{(k - 1)} = \Delta \tilde f_2^{(k - 1)} = \left\{ {\begin{array}{*{20}{c}}
{0,{\text{ for }}t < {t_{st}}{\text{ or }}t > {t_{sp}}}\\
{{\lambda _r}({\text{ }}t - {t_{st}}),{\text{  for }}{t_{st}} \le t \le {t_{sp}}}
\end{array}} \right.\\
{\text{where }}{\lambda _r} = 2\times 10^{-5},{t_{st}} = 0{\text{ s}},{t_{sp}} = 600{\text{ s}}
\end{array}
\end{equation}
\normalsize
The attack mimics a load reduction in both areas, increasing $\Delta f_1$ and $\Delta f_2$ while keeping the tie-line power fluctuating around zero (Fig. \ref{Fre_Ptie pre and post Coordinated Ramp Attacks}). This behavior resembles natural load variations, complicating detection. Basic rules (\ref{eq: basic rules}) fail to detect the attack, as the manipulated ACE displayed in the control room (the red trajectories in Fig. \ref{Fre_Ptie pre and post Coordinated Ramp Attacks}) remain within the bounds. Notably, if undetected, the true $f_1$ (the blue trajectory in Fig. \ref{Fre_Ptie pre and post Coordinated Ramp Attacks}) drops to the critical 59.5 Hz at 71.6 s, which may trigger load shedding and widespread outages. 
\begin{figure}[htbp]
\centering
\begin{subfigure}{0.16\textwidth}
  \centering
  \includegraphics[width=\textwidth]{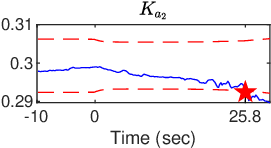}
  \caption{MLE}
\end{subfigure}%
\begin{subfigure}{0.16\textwidth}
  \centering
  \includegraphics[width=\textwidth]{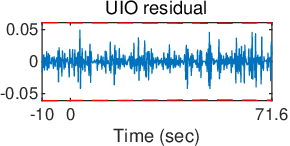}
  \caption{UIO \cite{Ameli2018}}
\end{subfigure}%
\begin{subfigure}{0.16\textwidth}
  \centering
  \includegraphics[width=\textwidth]{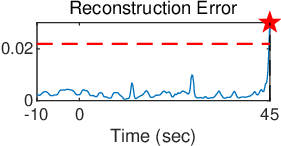}
  \caption{LSTM-AE \cite{Musleh2023}}
\end{subfigure}%
\caption{Comparative responses (detection indicators) to the coordinated ramp FDIA with $\lambda_r = 2\times 10^{-5}$ for both $\Delta f_1$ and $\Delta f_2$. Red dashed lines: thresholds; Red star: FDIA is detected.}
\label{coordinated_fre1_fre2_FDIA}
\end{figure}

In Fig. \ref{coordinated_fre1_fre2_FDIA}, the MLE detects the FDIA 25.8 s after initiation, while the LSTM-AE needs 45 s. In contrast, the UIO observer fails to identify the attack since the residuals are nearly the same before and during the whole FDIA period. 

%%%%%%%%%%%%%%%%
\subsection{3-Area System Configuration and Threshold Selection} \label{subsec:3area_config_and_threshold}

The 3-area system parameters are detailed in Table \ref{tab:3area_system}. The corresponding subvector is defined as ${x_{sub}} = {[\Delta {f_1},\Delta {f_2},\Delta {f_3},\Delta {P_{re{f_1}}},\Delta {P_{re{f_2}}},\Delta {P_{re{f_3}}},\Delta {P_{ti{e_{1 - 2}}}},\Delta {P_{ti{e_{2 - 3}}}}]^T}$. Similar to the 2-area configuration, the diffusion terms $\Gamma$ and mean load power deviations $\bm{\mu}_L$ are set to ensure relatively small Area Control Error (ACE) and tie-line power variations under normal operating conditions, satisfying \eqref{eq: basic rules}.

\begin{table}[H]
\centering
\caption{\textsc{Parameters of the 3-Area System}}
\label{tab:3area_system}
\resizebox{\linewidth}{!}
{
\begin{tabular}{|c|c|c|c|c|c|c|c|c|}
\hline
 & $H_i$ (s) & $D_i$ & $R_i$ (pu) & $T_{g_i}$ (s) & $T_{T_i}$ (s) & $K_{a_i}$ & $B_i$ & $K_{L_i}$ \\
\hline
Area 1 & 5 & 1.0 & 0.05 & 0.10 & 0.30 & 0.2 & 21.0 & 0.005 \\
\hline
Area 2 & 6 & 1.5 & 0.05 & 0.17 & 0.40 & 0.2 & 21.5 & 0.005 \\
\hline
Area 3 & 6 & 1.8 & 0.05 & 0.20 & 0.35 & 0.2 & 21.8 & 0.005 \\
\hline
Tie-line & \multicolumn{3}{c|}{$K_{tie_{1-2}} = 1.2478$} & \multicolumn{2}{c|}{$K_{tie_{2-3}} = 1.1498$} & \multicolumn{3}{c|}{Base Power = 1000 MVA} \\
\hline
\end{tabular}
}
\end{table}

With the system parameters established, we evaluate the detection performance and determine an appropriate threshold by simulating a ramp FDIA. The manipulated state vector ${\bm{\tilde x}}^{(k)}$ at time step $k$ is formulated as follows:
\begin{equation} \label{eq:ramp_attack_model_area1}
\begin{array}{l}
{{\bm{\tilde x}}^{(k)}} = {[\Delta \tilde f_1^{(k)}, {\bm{0}_{1 \times 16}}]^T}\\
\Delta \tilde f_1^{(k)} = \left\{ {\begin{array}{*{20}{c}}
{0,{\text{ for }}t < {t_{st}}{\text{ or }}t > {t_{sp}}}\\
{{\lambda _r}({\text{ }}t - {t_{st}}),{\text{ for }}{t_{st}} \le t \le {t_{sp}}}
\end{array}} \right.\\
{\text{where }}{\lambda _r} = 1\times 10^{-5},{t_{st}} = 0{\text{ s}},{t_{sp}} = 600{\text{ s}}
\end{array}
\end{equation}

\begin{figure}[htbp]
    \centering
    \includegraphics[width=0.3\textwidth]{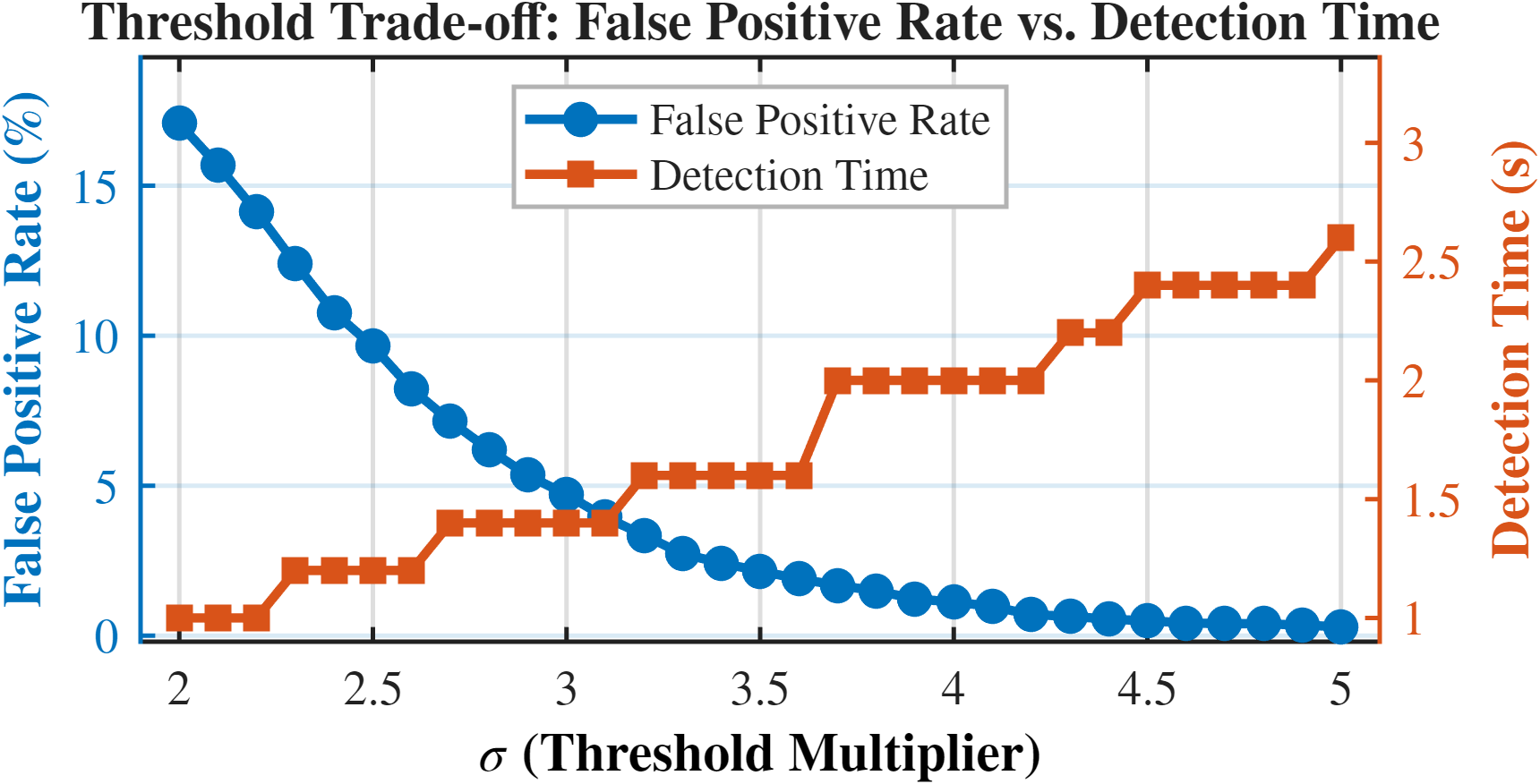}
    \caption{Threshold trade-off analysis: False Positive Rate vs. Detection Time across different $\sigma$ multipliers.}
    \label{fig:tradeoff}
\end{figure}

To explicitly justify the selection of the $4\sigma$ threshold under the attack conditions, a trade-off analysis between the False Positive Rate (FPR) and detection time was conducted across $\sigma \in [2, 5]$ (Fig.~\ref{fig:tradeoff}). As illustrated, lower threshold multipliers yield unacceptably high FPRs, which would trigger frequent false alarms and disrupt secondary frequency control. As the threshold increases, the FPR drops significantly while the detection time experiences a step-wise increase. The $4\sigma$ mark achieves an optimal balance, successfully suppressing the FPR to near-zero while maintaining a practically acceptable detection time of approximately 2 seconds. Increasing the threshold further yields diminishing returns, unnecessarily exacerbating the detection delay without meaningfully improving the already negligible FPR.

Furthermore, while this $4\sigma$ criterion was initially derived from the analysis of the 2-area system, our simulations demonstrate that it scales robustly to the 3-area system under the same $\lambda_r$ conditions. This confirms that the $4\sigma$ rule serves as a consistent and generalizable heuristic, providing the optimal sensitivity-specificity balance across different system configurations.

%%%%%%%%%%%%%%%%%

%%%%%%%%%%%%%%%%
\subsection{Concurrent FDIAs and Load Fluctuations}
\label{subsec:compounded_scenarios}

In practical power systems, malicious FDIAs may coincide with legitimate physical load fluctuations. To evaluate the robustness of the proposed method under such compounded conditions, we conducted a grid search of multi-area joint load changes ($\mu_{L_1}$, $\mu_{L_2}$, $\mu_{L_3}$) ranging from -0.2 pu to 0.2 pu with a 0.1 pu step, yielding 125 distinct extreme disturbance scenarios.

\begin{figure}[htbp]
\centering
\includegraphics[width=0.3\textwidth]{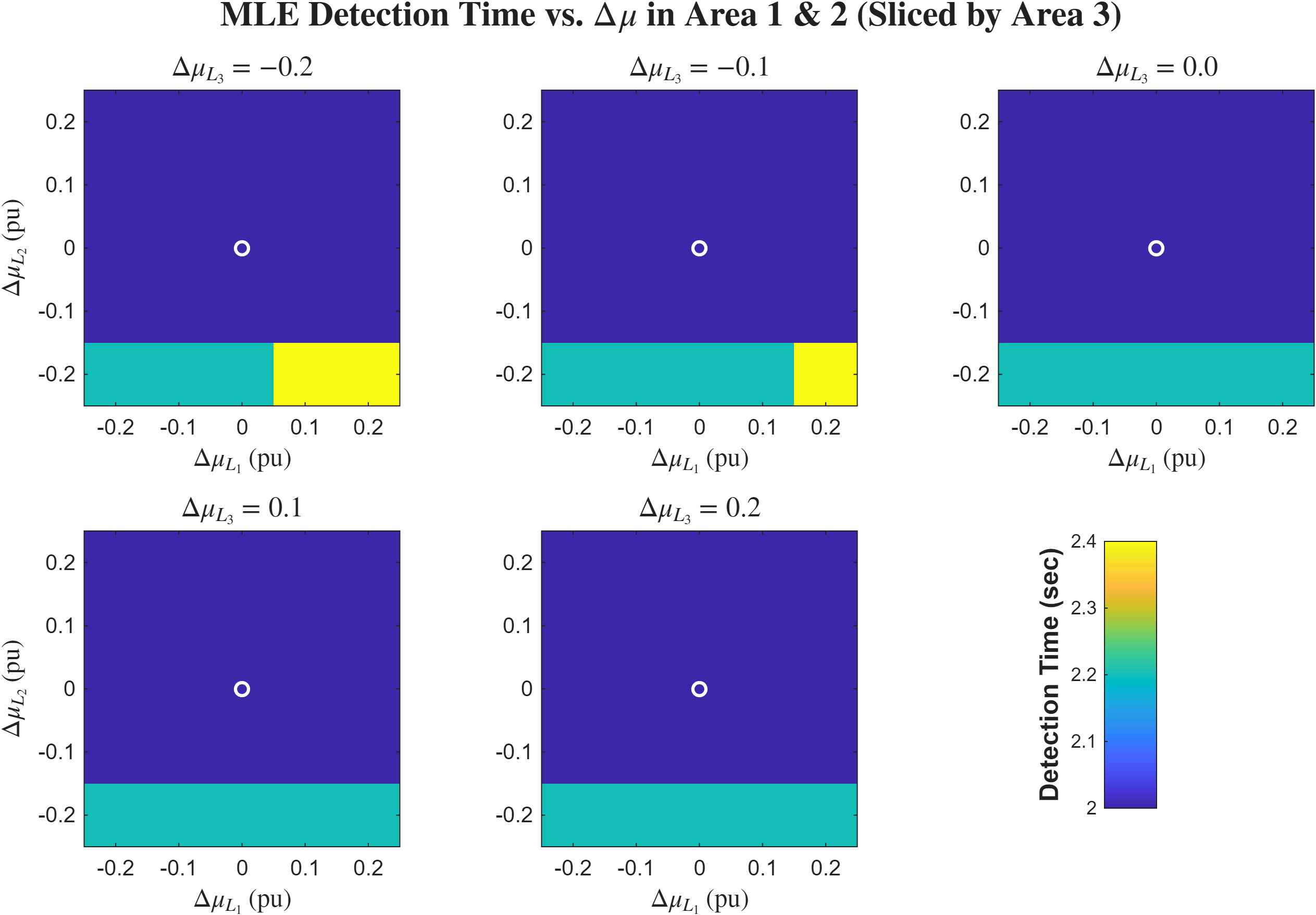}
\caption{Heatmaps of MLE detection time under simultaneous multi-area load jumps and FDIA. The 3D parameter space is visualized as 2D cross-sectional slices corresponding to different values of $\Delta\mu_{L_3}$. The white circles denote the nominal operating points ($\Delta\mu_{L_1}=0, \Delta\mu_{L_2}=0$).}
\label{fig:joint_load_jump}
\end{figure}

Fig. \ref{fig:joint_load_jump} visualizes the detection time when an FDIA occurs simultaneously with these load variations. The proposed method successfully isolates the FDIA component and detects the attack in \textit{all} 125 scenarios. The detection speed remains highly stable: it stays around 2.0 seconds for most cases (dark blue markers) and only slightly increases to a maximum of 2.4 seconds under the most severe simultaneous fluctuations.

Furthermore, we evaluated the False Positive Rate (FPR) across the same 125 scenarios in the absence of FDIA to ensure that normal load changes are not misclassified as attacks. The overall FPR is exceptionally low at 2.40\%, with only 3 extreme combinations triggering a false alarm: $(\mu_{L_1}, \mu_{L_2}, \mu_{L_3}) = (0.10, 0.10, 0.20)$ pu, $(0.00, 0.20, 0.20)$ pu, and $(0.10, 0.20, 0.20)$ pu. Notably, the trigger time for these rare false positives is significantly delayed (exactly 31.40 s) compared to the rapid detection of actual FDIAs (2.0--2.4 s). This substantial temporal gap provides an additional layer of security, effectively enabling system operators to distinguish extreme physical load jumps from malicious cyberattacks.
%%%%%%%%%%%%%%%%
\subsection{Comparisons with Kalman in the 3-Area System} 
To demonstrate the advantages of the proposed method, we implemented a discrete-time Kalman Filter (KF) cascaded with a Cumulative Sum (CUSUM) detector for the 3-area AGC system. 
By extracting the locally observable subsystem, the discrete-time state transition and measurement models are formulated as:
\begin{align}
\label{eq: KF_system_x}
\bm{x}^{(k)} &= e^{A_{sub}\Delta t}\bm{x}^{(k - 1)} + (I - e^{A_{sub}\Delta t})\bm{\mu}_t + S\bm{\xi}_t \\
\label{eq: KF_system_y}
\bm{y}^{(k)} &= \bm{x}^{(k)} + \bm{v}^{(k)}
\end{align}
where the state vector $\bm{x}^{(k)}$ encompasses the observable frequencies, tie-line powers, and generation commands of the 3-area system. Based on this model, the standard KF recursively estimates the states and computes the measurement residual (innovation) $\tilde{\bm{y}}^{(k)}$. The prediction and update phases are compactly given by:
\begin{align}
\label{eq: KF_predict}
&\hat{\bm{x}}^{(k|k - 1)} = e^{A_{sub}\Delta t}\hat{\bm{x}}^{(k - 1|k - 1)} + (I - e^{A_{sub}\Delta t})\bm{\mu}_t \\
\label{eq: KF_cov_predict}
&P^{(k|k - 1)} = e^{A_{sub}\Delta t}P^{(k - 1|k - 1)}(e^{A_{sub}\Delta t})^T + Q_{kf} \\
\label{eq: KF_residual_S}
&\tilde{\bm{y}}^{(k)} = \bm{y}^{(k)} - \hat{\bm{x}}^{(k|k - 1)}, \quad S_{kf}^{(k)} = P^{(k|k - 1)} + R_{kf} \\
\label{eq: KF_gain_update}
&K_{kf}^{(k)} = P^{(k|k - 1)}(S_{kf}^{(k)})^{- 1}, \quad \hat{\bm{x}}^{(k|k)} = \hat{\bm{x}}^{(k|k - 1)} + K_{kf}^{(k)}\tilde{\bm{y}}^{(k)} \\
\label{eq: KF_cov_update}
&P^{(k|k)} = (I - K_{kf}^{(k)})P^{(k|k - 1)}
\end{align}

To detect anomalies using the KF outputs, a CUSUM test is applied to the normalized residual. Under normal operating conditions, the Mahalanobis distance of the residual, defined as $(\tilde{\bm{y}}^{(k)})^T (S_{kf}^{(k)})^{-1} \tilde{\bm{y}}^{(k)}$, approximately follows a $\chi^2$ distribution. The non-parametric CUSUM statistic $g^{(k)}$ recursively accumulates deviations that exceed a predefined drift parameter:
\begin{equation} \label{eq: CUSUM_update}
g^{(k)} = \max(0, g^{(k - 1)} + (\tilde{\bm{y}}^{(k)})^T (S_{kf}^{(k)})^{-1} \tilde{\bm{y}}^{(k)} - \nu_{kf})
\end{equation}
An attack alarm is triggered whenever the accumulated statistic $g^{(k)}$ exceeds a detection threshold $\tau_{kf}$. 

\begin{figure}[htbp]
  \centering
  \begin{subfigure}{0.16\textwidth}
    \centering
    \includegraphics[width=\textwidth]{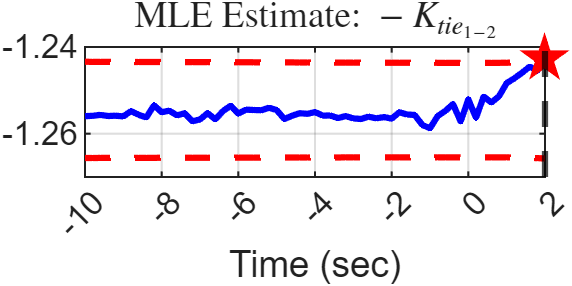}
    \caption{MLE}
    \label{fig:mle_detection}
  \end{subfigure}\hfill
  \begin{subfigure}{0.16\textwidth} 
    \centering
    \includegraphics[width=\textwidth]{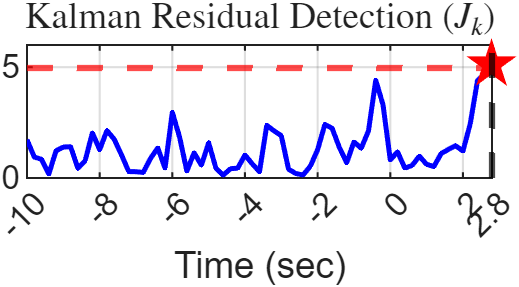}
    \caption{Kalman Residual}
    \label{fig:kf_residual}
  \end{subfigure}\hfill
  \begin{subfigure}{0.16\textwidth} 
    \centering
    \includegraphics[width=\textwidth]{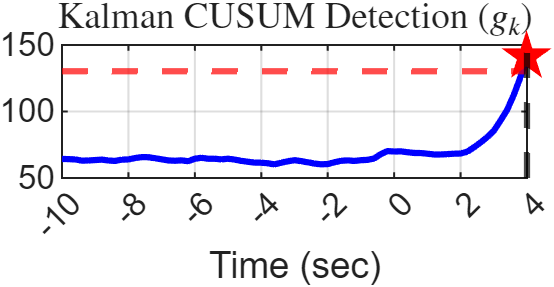}
    \caption{Kalman CUSUM}
    \label{fig:kf_cusum}
  \end{subfigure}
  \caption{Comparative detection responses to the ramp FDIA for $\Delta f_1$. Red dashed lines represent the detection thresholds, and red stars indicate the exact moment the FDIA is detected.}
  \label{fig:comparison_mle_kf}
\end{figure}

As illustrated in Fig. \ref{fig:comparison_mle_kf}, the three subfigures present the detection trajectories of the MLE method, the Kalman Residual ($J^{(k)}$), and the Kalman CUSUM ($g^{(k)}$) under the same ramp FDIA targeting $\Delta f_1$. The results clearly indicate that the MLE method achieves a faster detection response than both KF-based approaches. This performance gap fundamentally stems from the practical difficulty of accurately estimating the measurement noise covariance matrix ($R_{kf}$) in real-world environments. In a practical power system, true measurement noise cannot be perfectly isolated from operational data. Consequently, the empirical $R_{kf}$ used in the KF inevitably contains mismatches, which degrades its tracking accuracy and delays the accumulation of both the residual and CUSUM statistics. While the KF could theoretically outperform MLE under idealized conditions—assuming perfect prior knowledge where $R_{kf}$ matches the true noise distribution exactly—such scenarios are unattainable in practice. Therefore, the data-driven MLE approach demonstrates superior robustness and a more rapid detection capability for realistic applications.
%%%%%%%%%%%%%%%%

%%%%%%%%%%

%%%%%%%%%%%%%%%%
\subsection{Robustness Analysis and Failure Boundaries of the Estimation Method}

In this section, we analyze the approximation error introduced by setting $A_2 = 0$ and define the exact failure boundaries of the proposed MLE method. The MLE approach is inherently robust under normal operating conditions because the dynamics of the unmeasured states are typically well-damped. This adequate damping prevents the unmeasured states from persistently exciting the observable states. To quantify this robustness and identify where the method fails, we conduct a sensitivity analysis by progressively reducing the system's equivalent inertia. Specifically, we introduce an inertia multiplier ($M_{scale}$) that directly scales the nominal inertia constants ($H_i$) of all areas detailed in Table \ref{tab:3area_system} (i.e., $H_{i,scaled} = M_{scale} \times H_i$).

\begin{figure}[htbp]
    \centering
    \includegraphics[width=0.4\textwidth]{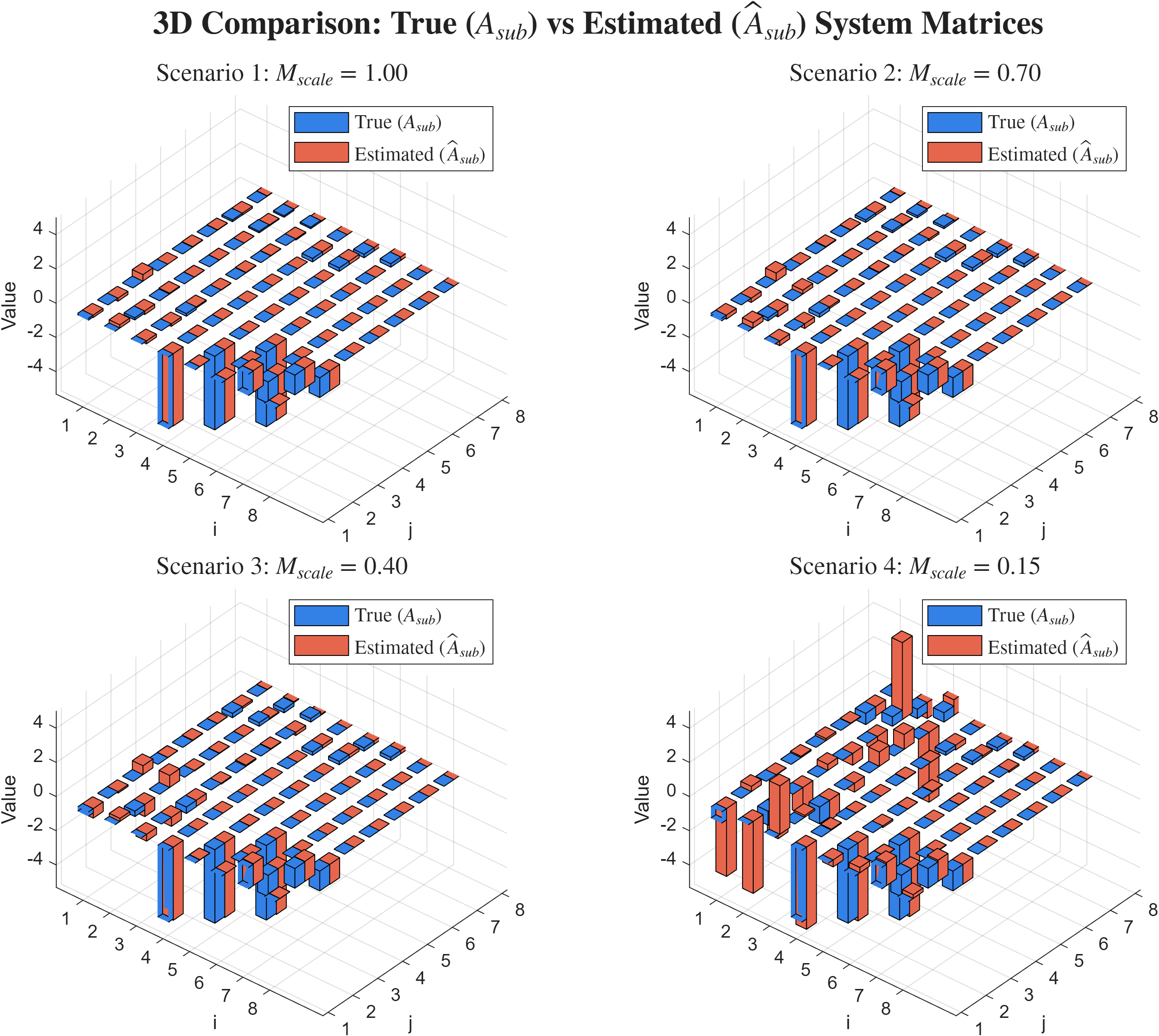}
    \caption{3D Comparison: True (blue) vs Estimated (red) System Matrices ($A_{sub}$) under varying inertia. The estimation is highly robust under Scenarios 1-3, but fails in Scenario 4.}
    \label{fig:3d_comparison}
\end{figure}

\textbf{Error Quantification:} We first compare the true subsystem matrix ($A_{sub}$) with the estimated matrix ($\hat{A}_{sub}$). As illustrated in Fig. \ref{fig:3d_comparison}, the 3D bar charts display the true matrix elements in blue and the estimated elements in red across four inertia scenarios. Under normal to moderately stressed conditions (Scenarios 1-3, $M_{scale} \ge 0.40$), the blue and red bars align almost perfectly. This visual confirmation demonstrates that the approximation error is negligible, further validating the robustness of the MLE method when sufficient system inertia is maintained.

However, the parameter estimation severely diverges under extremely low inertia (Scenario 4, $M_{scale} = 0.15$), where large red spikes emerge, highlighting massive estimation errors.

\begin{figure}[htbp]
    \centering
    \includegraphics[width=0.3\textwidth]{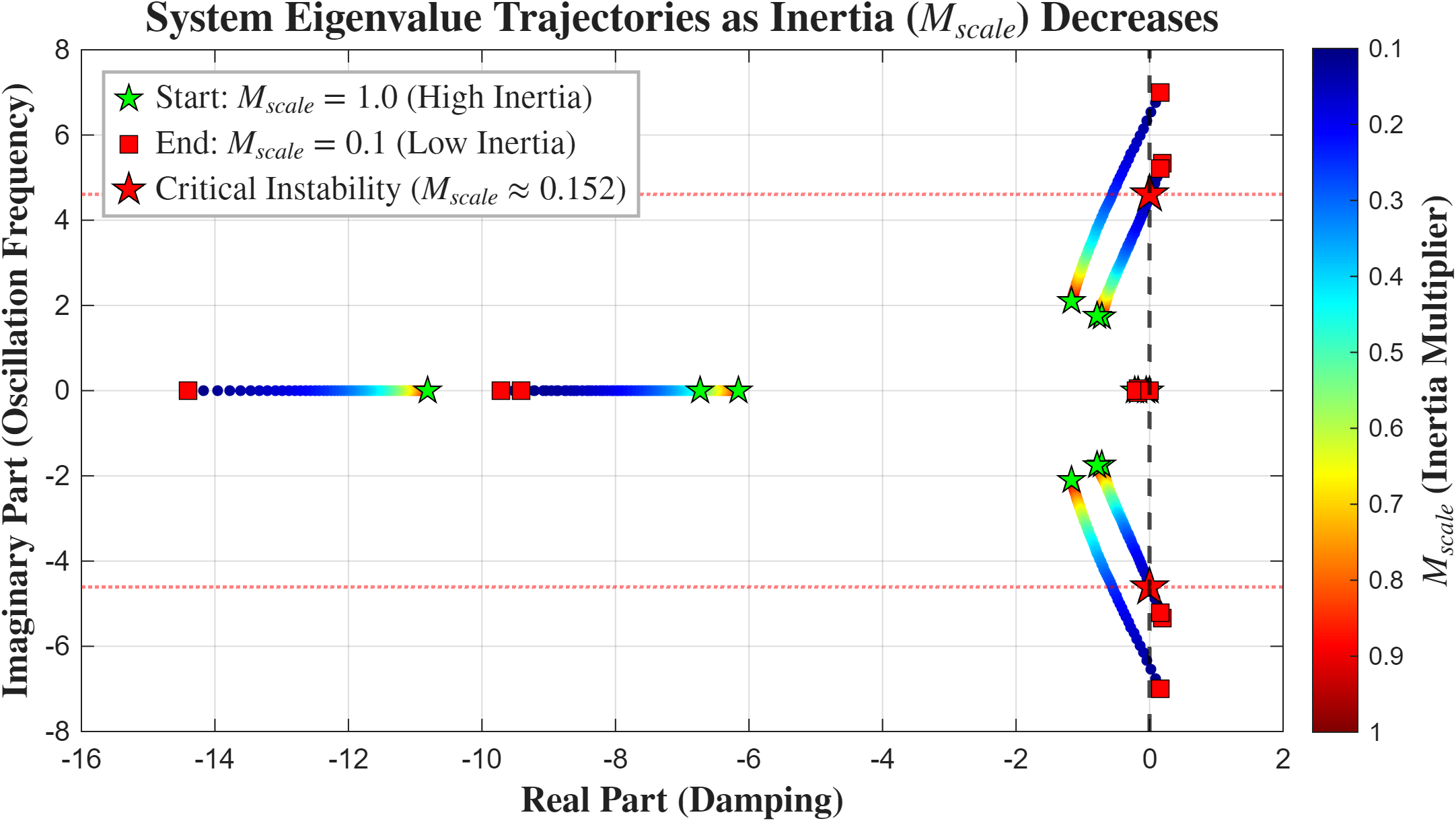}
    \caption{System Eigenvalue Trajectories tracing the shift towards instability. The MLE method fails at the critical boundary where eigenvalues approach the imaginary axis.}
    \label{fig:eigenvalue_trajectories}
\end{figure}

\textbf{Failure Boundary and Root Cause:} To mathematically explain the root cause of this divergence, Fig. \ref{fig:eigenvalue_trajectories} plots the system's eigenvalue trajectories. The plot tracks the eigenvalues moving from the left half-plane towards the imaginary axis as the inertia multiplier drops from $M_{scale} = 1.0$ (green stars) to $M_{scale} = 0.1$ (red squares).

The estimation fails precisely at $M_{scale} \approx 0.152$ (red stars), where dominant complex eigenvalues approach the imaginary axis and eliminate physical damping. The resulting undamped oscillations in unmeasured states violate the OU model's mean-reverting property, invalidating the $A_2 = 0$ decoupling approximation. Practically, however, this failure only occurs near dynamic instability, where standard BDD alarms or protection schemes would inherently trigger before cyberattack detection is needed. Thus, the proposed method remains highly reliable within standard operational bounds.

%%%%%%%%%%%%%%%%
\subsection{Detecting ACE Attacks in 3-Area Systems}
To thoroughly evaluate the proposed detection mechanism, we test another ``smart'' negative-compensation FDIA targeting the ACE, as proposed in \cite{Ameli2018}. To enhance its stealthiness and evade conventional detection, this sophisticated attack scheme is implemented as a progressive ramp attack rather than an abrupt step change. When load variations produce non-zero $\Delta P_{L_i}$ and $ACE_i$ values, the attacker intercepts and manipulates the critical measurements before they reach the control center. This is achieved by gradually applying a time-varying scaling factor, $\alpha^{(k)}$, to both the frequency deviation $\Delta f_i^{(k)}$ and all associated tie-line power measurements $\Delta P_{tie_{i-j}}^{(k)}$ ($\forall j \in \Omega_i$).

This coordinated manipulation leads to the following altered ACE signal. To avoid triggering threshold-based alarms instantly, the scaling factor $\alpha^{(k)}$ is designed to decrease linearly from $1$ (normal operation) to a final negative-compensation value $\alpha$ over a specific transition period. The dynamic evolution of the manipulated signal and $\alpha^{(k)}$ is defined as:
\begin{equation}
\begin{aligned}
    &ACE_1^{(k)} + \Delta \widetilde{ACE}_1^{(k)} = \alpha^{(k)} ACE_1^{(k)} \\
    &= {B_1}(\Delta f_1^{(k)} + \Delta \tilde{f}_1^{(k)}) + \Delta P_{tie_{1-2}}^{(k)} + \Delta \tilde{P}_{tie_{1-2}}^{(k)} \\
    \alpha^{(k)} &= \begin{cases}
        1, & \text{for } t < t_{st} \\
        1 + (\alpha - 1)\frac{t - t_{st}}{t_{sp} - t_{st}}, & \text{for } t_{st} \le t \le t_{sp} \\
        \alpha, & \text{for } t > t_{sp}
    \end{cases}
\end{aligned}
\end{equation}
where $\alpha = -1$, $t_{st} = 0 \text{ s}$, and $t_{sp} = 600 \text{ s}$.

In our simulation scenario, we analyze a negative-compensation FDIA that progressively targets the $ACE_1$ signal with a final scaling factor of $\alpha = -1$. This attack deliberately inverts the authentic ACE signal over time. If sustained, such an FDIA creates a persistent and growing mismatch between load and generation, ultimately leading to severe frequency deviations in the physical power system.

\begin{figure}[htbp]
    \centering
    \begin{subfigure}[b]{0.5\textwidth}
        \centering
        \includegraphics[width=1\linewidth]{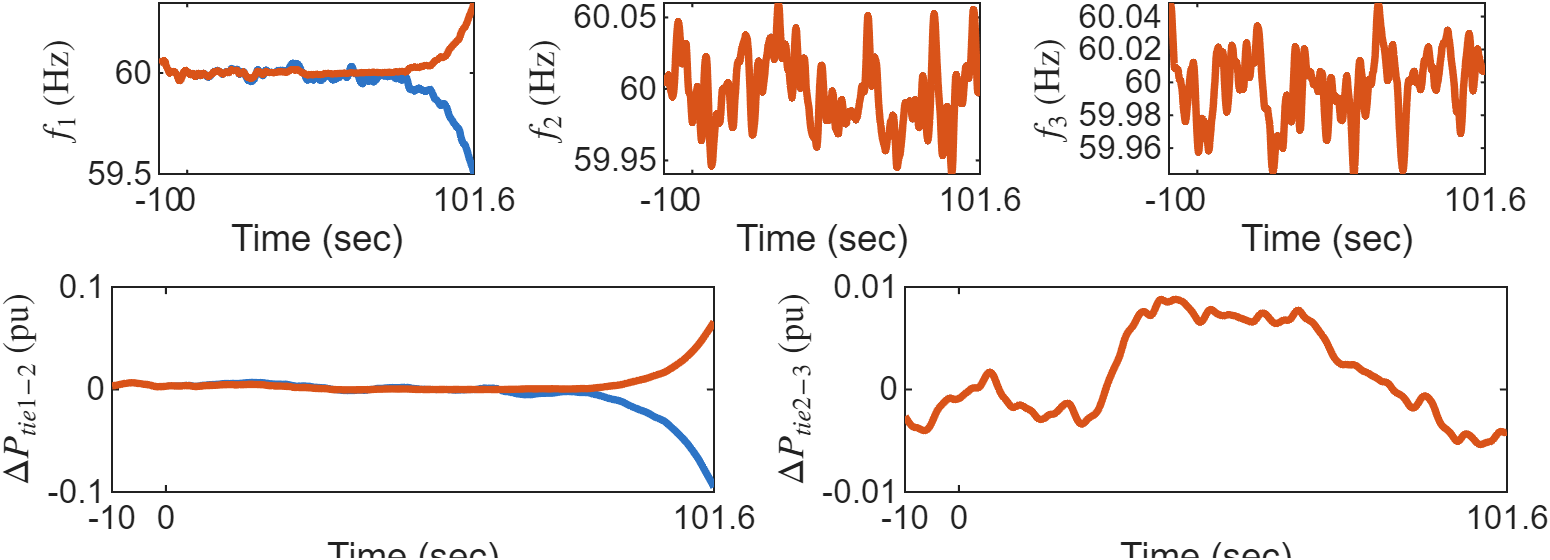}
        \caption{Physical vs. virtual system responses}
        \label{fig:ACE_FDIA_responses}
    \end{subfigure}
    \begin{subfigure}[b]{0.5\textwidth} 
        \centering
        \includegraphics[width=0.32\textwidth]{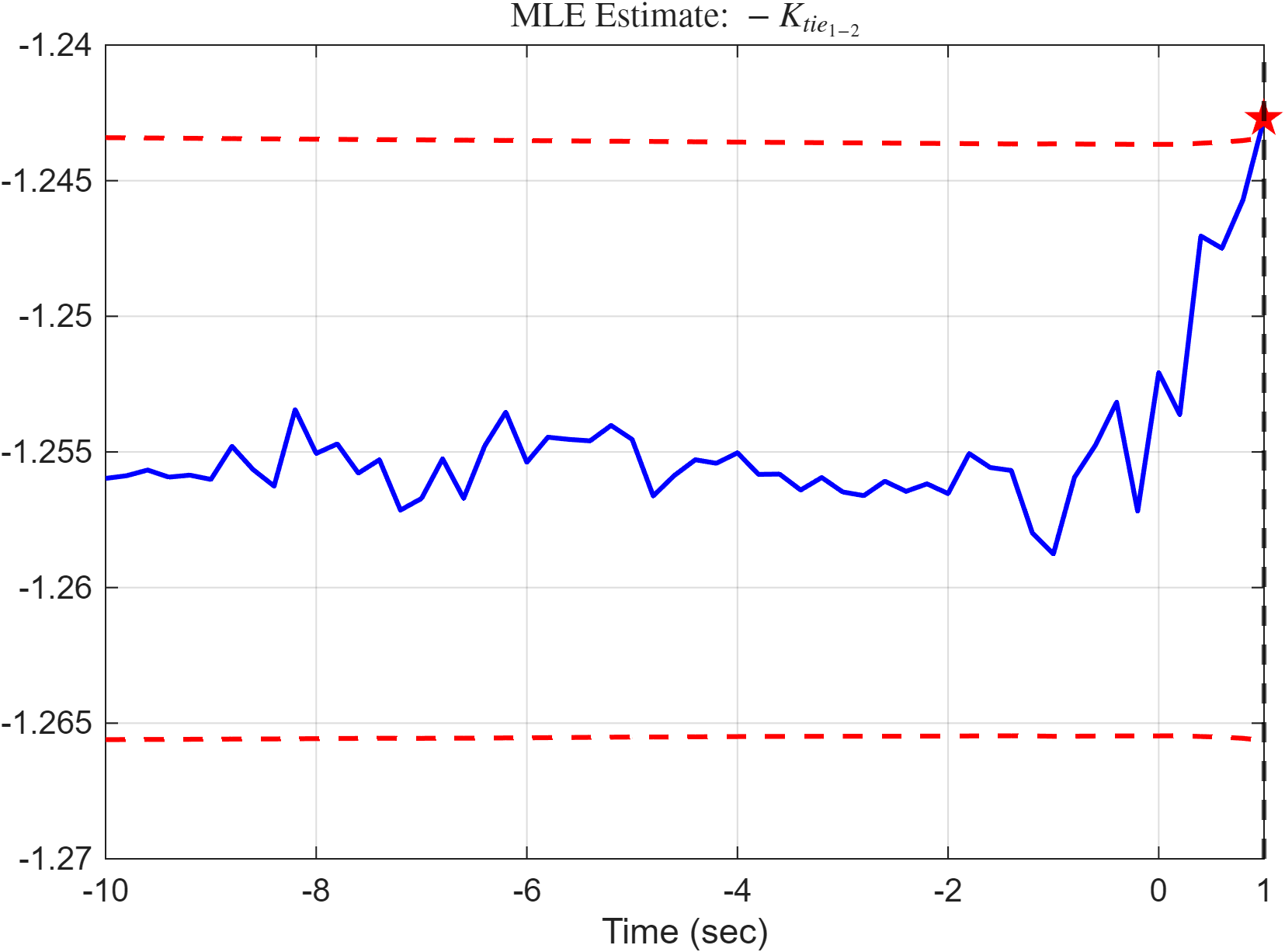} 
        \caption{MLE detection indicator}
        \label{fig:ACE_FDIA_Detection_sub}
    \end{subfigure}
    \caption{(a) System responses and (b) MLE detection indicator before and during the progressive negative-compensation FDIA targeting $ACE_1$. Blue: True physical trajectories, where the attack drives $f_1$ to 59.5 Hz; Red: Manipulated virtual trajectories in the control center, appearing stable to avoid alarms; Red star: Successful detection point by the proposed MLE method.}
    \label{fig:Combined_ACE_Figure}
\end{figure}

Fig. \ref{fig:ACE_FDIA_responses} illustrates the severe consequences of this stealthy manipulation. If undetected, the ramped FDIA drives the true physical frequency $f_1$ (the blue trajectory) to the critical termination limit of 59.5 Hz while inducing dangerous system-wide oscillations. Concurrently, the basic detection rule defined in (\ref{eq: basic rules}) fails to identify the anomaly, as the manipulated virtual measurements (the red trajectories) scale smoothly and deliberately remain within standard operational bounds.

To counter this, Fig. \ref{fig:ACE_FDIA_Detection_sub} demonstrates the effectiveness of the proposed MLE detector. Despite the attack's smooth transition and visually normal operating data, the MLE method successfully detects the progressive negative-compensation FDIA during its early ramp phase. Specifically, the anomaly is identified (indicated by the red star) well before the physical system frequency can degrade to critical limits.
%%%%%%%%%%%%%%%%%%%%%%%%%%%%%%%%%

%%%%%%%%%%%%%%%%

%%%%%%%%%%%%%%%%
\section{Conclusions}
This paper proposes a robust framework for detecting stealthy FDIAs in AGC systems by leveraging the MLE of a drifted multivariate OU process. This approach estimates the system state matrix directly from accessible measurements, entirely circumventing the need for real-time load data or detailed system parameters. By monitoring deviations in estimated AGC parameters, the proposed algorithm can detect FDIAs.

Extensive evaluations demonstrate that the proposed MLE detector significantly outperforms traditional UIO, LSTM-AE, and KF-based approaches in accuracy and detection speed. It rapidly identifies coordinated attacks that evade UIO, avoids the severe load-variation sensitivities of LSTM-AE, and overcomes the practical limitations of KF regarding measurement noise covariance estimation. Supported by a dynamic threshold mechanism, this probabilistic framework maintains a low FPR even under simultaneous multi-area load jumps, providing a highly reliable tool to secure cyber-physical power grids.

However, the online MLE involves high computational overhead due to matrix exponentials, covariance inversion, and sliding windows, posing challenges for direct implementation on resource-constrained edge devices. Future work will explore lightweight approximations and edge-cloud collaborative architectures to facilitate practical real-time deployment.

%%%%%%%%%%%%%%%%%%%%%%%%%%%%%%%%%
\appendix
\label{appendixA}
We first review a few properties in matrix calculus, which will be used to prove  \textbf{Theorem 1}. Let $y$ be a scalar function of an $N \times N$ matrix $X$ or a $N \times 1$ vector $\bm{x}$. Then, using numerator-layout notation, we define:
\small
\begin{equation}
\begin{array}{l}
\partial y/\partial X = \left[ {\begin{array}{*{20}{c}}
{\partial y/\partial {X_{11}}}&{...}&{\partial y/\partial {X_{N1}}}\\
{...}&{...}&{...}\\
{\partial y/\partial {X_{1N}}}&{...}&{\partial y/\partial {X_{NN}}}
\end{array}} \right]\\
\partial y/\partial \bm{x} = \left[ {\begin{array}{*{20}{c}}
{\partial y/\partial {x_1}}&{...}&{\partial y/\partial {x_N}}
\end{array}} \right]
\end{array}
\end{equation}
\normalsize
Let $\bm{a},\bm{b}$ be constant $N\times 1$ column vectors, and $C$, $F$ be constant $N\times N$ matrices. The following properties \cite{Crowder2019} apply:
\small
\begin{flalign}
&\scalebox{0.95}{$\partial {(X\bm{a} + \bm{b})^T}C(X\bm{a} + \bm{b})/\partial X = {\left( {\left( {C + {C^T}} \right)(X\bm{a} + \bm{b}){\bm{a}^T}} \right)^T}$} && \label{eq: Preliminaries_1} \\
&\scalebox{0.95}{$\partial {\bm{a}^T}X\bm{b}/\partial X = \bm{b}{\bm{a}^T}$} && \label{eq: Preliminaries_2} \\
&\scalebox{0.95}{$\partial \log \left| X \right|/\partial X = {X^{ - 1}}$} && \label{eq: Preliminaries_3} \\
&\scalebox{0.95}{$\log \left| {{X^{ - 1}}} \right| =  -\log \left| X \right|$} && \label{eq: Preliminaries_4} \\
&\scalebox{0.95}{$\partial {(F\bm{x} + \bm{b})^T}C(F\bm{x} + \bm{b})/\partial \bm{x} = {(F\bm{x} + \bm{b})^T}({C^T} + C)F$} && \label{eq: Preliminaries_5} 
\end{flalign}
\normalsize
\footnotesize 
\begin{align}         
&\begin{aligned}\label{eq: derivative of A proof}  
  &\partial L(A,\bm{\mu },{\Sigma ^{ - 1}})/\partial {e^{A\Delta t}} = 0 \\
  &\Rightarrow ({\Sigma ^{ - 1}} + {({\Sigma ^{ - 1}})^T}) \\
  &\qquad \times \sum\limits_{k = 1}^M {({\bm{x}^{(k)}} - \bm{\mu } - {e^{A\Delta t}}({\bm{x}^{(k - 1)}} - \bm{\mu }))} {({\bm{x}^{(k - 1)}} - \bm{\mu })^T} = 0 \\
  &\Rightarrow {e^{A\Delta t}}\sum\limits_{k = 1}^M {\left( {({\bm{x}^{(k - 1)}} - \bm{\mu }){{({\bm{x}^{(k - 1)}} - \bm{\mu })}^T}} \right)} \\
  &\qquad =\sum\limits_{k = 1}^M {\left( {({\bm{x}^{(k)}} - \bm{\mu }){{({\bm{x}^{(k - 1)}} - \bm{\mu })}^T}} \right)} 
\end{aligned}  \\
&\begin{aligned}\label{eq: derivative of mu proof}
  &\partial L(A,\bm{\mu },{\Sigma ^{ - 1}})/\partial \bm{\mu } = 0 \\
  &\Rightarrow \sum\limits_{k = 1}^M {{{(({e^{A\Delta t}} - I)\bm{\mu } + {\bm{x}^{(k)}} - {e^{A\Delta t}}{\bm{x}^{(k - 1)}})}^T}} \\
  &\qquad \times ({\Sigma ^{ - 1}} + {({\Sigma ^{ - 1}})^T})({e^{A\Delta t}} - I) = 0 \\
  &\Rightarrow (I - {e^{A\Delta t}})\bm{\mu}=\frac{1}{M}\sum\limits_{k = 1}^M {({\bm{x}^{(k)}} - {e^{A\Delta t}}{\bm{x}^{(k - 1)}})}  
\end{aligned}  \\
&\begin{aligned}\label{eq: derivative of Sigma proof}
  &\partial L(A,\bm{\mu },{\Sigma ^{ - 1}})/\partial {\Sigma ^{ - 1}} = 0 \\
  &\Rightarrow M{({\Sigma ^{ - 1}})^{ - 1}} = \sum\limits_{k = 1}^M {(({\bm{x}^{(k)}} - \bm{\mu } - {e^{A\Delta t}}({\bm{x}^{(k - 1)}} - \bm{\mu }))} \\
  &\qquad \times {({\bm{x}^{(k)}} - \bm{\mu } - {e^{A\Delta t}}({\bm{x}^{(k - 1)}} - \bm{\mu }))^T}) \\
\end{aligned} 
\end{align}
\normalsize
where $\left| X \right|$ is the determinant of the $X$. The detailed derivations can be found in \cite{Crowder2019}. In addition, according to the symmetry of the covariance matrix $\Sigma$, we have ${\Sigma ^{ - 1}} = {({\Sigma ^{ - 1}})^T}$. Though the matrix $S{S^T}$ from \eqref{eq: AGC_system_compact} is theoretically singular, making $\Sigma$ symmetric but non-invertible, measurement noise in practice ensures the estimated $\Sigma$ is almost surely invertible. 

Consider the log-likelihood function $L(A,\bm{\mu },{\Sigma ^{ - 1}})$ in (\ref{eq: multivariate likelihood}), if we use the properties described in \eqref{eq: Preliminaries_1}-\eqref{eq: Preliminaries_5}, we can solve for $A$, $\bm{\mu }$ and $\Sigma$ from the condition $\partial L(A,\bm{\mu },{\Sigma ^{ - 1}})/\partial {e^{ A\Delta t}}=0$, $\partial L(A,\bm{\mu },{\Sigma ^{ - 1}})/\partial \bm{\mu } = 0$, $\partial L(A,\bm{\mu },{\Sigma ^{ - 1}})/\partial {\Sigma ^{ - 1}}=0$, i.e., the necessary condition for the maximum of $L(A,\bm{\mu },{\Sigma ^{ - 1}})$. 

Using \eqref{eq: Preliminaries_1} for $\partial L(A,\bm{\mu },{\Sigma ^{ - 1}})/\partial {e^{ A\Delta t}}=0$, using \eqref{eq: Preliminaries_5} for $\partial L(A,\bm{\mu },{\Sigma ^{ - 1}})/\partial {\bm{\mu }} = 0$, using \eqref{eq: Preliminaries_2}-\eqref{eq: Preliminaries_4} for $\partial L(A,\bm{\mu },{\Sigma ^{ - 1}})/\partial {\Sigma ^{ - 1}} = 0$, then a set of simultaneous equations \eqref{eq: derivative of A proof}, \eqref{eq: derivative of mu proof}, and \eqref{eq: derivative of Sigma proof} can be obtained, which completes the proof. 
%%%%%%%%%%%%%%%%%%%%%%%%%%
\bibliographystyle{IEEEtran}
\bibliography{My_Collection}
%%%%%%%%%%%%%%%%
\end{document}